\documentclass[preprint,amsmath,amssymb,aps]{revtex4-1}
\selectfont

\usepackage[colorlinks=true,citecolor=blue,urlcolor=blue]{hyperref}
\usepackage{graphicx}       
\usepackage{dcolumn}        
\usepackage{bm}             
\usepackage{amssymb}
\usepackage{amstext}
\usepackage{tensor}
\usepackage[utf8]{inputenc}
\usepackage{color}
\usepackage[usenames,dvipsnames,svgnames,table]{xcolor}
\usepackage{multirow}

\newcommand{\beq}{\begin{equation}}
\newcommand{\eeq}{\end{equation}}
\newcommand{\bnab}{\boldsymbol{\nabla}}

\begin{document}

 \title{Stable photon orbits in stationary axisymmetric electrovacuum spacetimes}

\author{Sam R.~Dolan}
\email{s.dolan@sheffield.ac.uk}
\affiliation{Consortium for Fundamental Physics, School of Mathematics and Statistics, University of Sheffield, Hicks Building, Hounsfield Road, Sheffield S3 7RH, United Kingdom.}

\author{Jake O.~Shipley}
\email{joshipley1@sheffield.ac.uk}
\affiliation{Consortium for Fundamental Physics, School of Mathematics and Statistics, University of Sheffield, Hicks Building, Hounsfield Road, Sheffield S3 7RH, United Kingdom.}

\date{\today}

\begin{abstract}
We investigate the existence and phenomenology of stable photon orbits (SPOs) in stationary axisymmetric electrovacuum spacetimes in four dimensions. First, we review the classification of equatorial circular photon orbits on Kerr--Newman spacetimes in the charge-spin plane. 
Second, using a Hamiltonian formulation, we show that Reissner--Nordstr\"om di-holes (a family encompassing the Majumdar--Papapetrou and Weyl--Bach special cases) admit SPOs, in a certain parameter regime that we investigate. Third, we explore the transition from order to chaos for typical SPOs bounded within a toroidal region around a di-hole, via a selection of Poincar\'e sections. Finally, for general axisymmetric stationary spacetimes, we show that the Einstein--Maxwell field equations allow for the existence of SPOs in \emph{electro}vacuum; but not in pure vacuum. 
\end{abstract}

\maketitle

\section{Introduction\label{sec:intro}}

On 14th September 2015, gravitational waves from a compact-binary coalescence were observed for the first time, in both detectors of the aLIGO experiment \cite{Abbott:2016blz}; and another signal was confirmed later in the first observing run \cite{Abbott:2016nmj, TheLIGOScientific:2016pea}. The `chirp' profiles of events GW150914 and GW151226 have the characteristic inspiral, merger and ringdown phases anticipated in a binary black hole merger \cite{TheLIGOScientific:2016qqj,Buonanno:2006ui}.

The ringdown phase is strong evidence that the merged body possesses a `light-ring'~\cite{Cardoso:2016rao}, that is, a family of \emph{unstable} photon orbits \cite{Teo:2003, Hod:2012ax}. Heuristically, the orbital frequencies and Lyapunov exponents of the photon orbits are linked to the frequency and decay rate of the ringdown phase \cite{Goebel:1972, Mashhoon:1985cya, Cardoso:2008bp, Dolan:2010wr, Yang:2012he}. A light-ring will generate as-yet-unobserved phenomena, such as multiple lensing images around black hole shadows \cite{Perlick:2004tq}; diffraction effects such as glories and orbiting \cite{Crispino:2009xt}; and a characteristic spectrum of quasinormal modes \cite{Konoplya:2011qq}.

Taken in isolation, the first detections do not rule out the possibility that the merger end-product possesses a light-ring \emph{but not} an event horizon \cite{Cardoso:2016rao}. However, in horizonless scenarios (e.g.~ultra-compact boson stars \cite{Liebling:2012fv}, gravastars \cite{Chirenti:2016hzd} or wormholes \cite{Damour:2007ap}), the outer unstable photon orbits are generically accompanied by inner \emph{stable} photon orbits \cite{Cardoso:2014sna}. Stable photon orbits (SPOs) are associated with distinct phenomenological features, such as (i) trapping and storage of electromagnetic energy in bound regions, allowing various instabilities to flourish (e.g.~fragmentation and collapse \cite{Cardoso:2014sna}, and/or an ergoregion instability \cite{Friedman:1978, Pani:2010jz}); (ii) slow logarithmic decay of perturbations with time \cite{Keir:2014oka}, dominating over power-law (Price) decay \cite{Casals:2015nja}; (iii) distinctive chaotic features in black hole shadows \cite{Shipley:2016omi, Cunha:2015yba}; (iv) internal reflection, and thus a modified late-time ringdown \cite{Cardoso:2016rao}. The latter possibility will surely be tested and constrained by future gravitational-wave detections.

Are SPOs relevant {\it only} in exotic horizonless scenarios? No. Several strands of evidence hint at a wider role. It has been known for decades that SPOs exist, in principle, inside the inner horizons of Kerr--Newman black holes and around naked singularities \cite{Liang:1974, Calvani:1980is, Stuchlik:1981, Balek:1989, Pugliese:2013zma, Dokuchaev:2011wm, Ulbricht:2015vwa, Khoo:2016xqv}; and around black holes or solitons with a cosmological constant \cite{Stuchlik:2002,Grenzebach:2014fha}. Recently, SPOs have been revealed in the exterior regions of, first, higher-dimensional black rings \cite{Igata:2013be} and black holes \cite{Igata:2014xca}, second, Majumdar--Papapetrou di-holes in four dimensions \cite{Shipley:2016omi} (see also \cite{Wunsch:2013st}). As these di-holes may be viewed as static, axisymmetric toy models for binary black hole systems, the possibility of SPOs arising in nature cannot be hastily dismissed.

Here we explore SPOs in stationary axisymmetric spacetimes in the four-dimensional electrovacuum context. The article comes in four parts: Sec.~\ref{sec:KN}, a review of the classification of equatorial circular photon orbits in Kerr--Newman spacetimes; Sec.~\ref{sec:dihole}, highlighting the existence of SPOs in static di-hole systems; Sec.~\ref{sec:poincare}, exploring the structure of SPOs through their Poincar\'e sections; and Sec.~\ref{sec:existence}, a key result on the existence of SPOs under the rather general assumptions of stationarity, axisymmetry and electrovacuum. We conclude in Sec.~\ref{sec:discussion} with a discussion of physical implications.

\section{Stable photon orbits}

\subsection{Equatorial circular photon orbits in Kerr--Newman spacetimes\label{sec:KN}}
Black hole uniqueness theorems \cite{Robinson:1975bv, Mazur:1982db} support the conjecture that the final product of gravitational collapse in asymptotically flat electrovacuum is a Kerr--Newman (KN) black hole, described by just three parameters: mass $M$, charge ratio $q = Q/M$ and spin ratio $a = J/M^2$. The KN family may be extended beyond the black hole case to include naked singularities with $a^2 + q^2 > 1$. The geodesic equations in KN spacetimes are separable, thanks to the existence of the Carter constant \cite{Carter:1968rr}. The problem of classifying the equatorial circular photon orbits (ECPOs) of non-zero energy reduces to classifying the repeated roots of a certain quartic,
$
\mathcal{R}(u) = 1 - (b^2-a^2) u^2 + (a-b)^2 (2 u^3 - q^2 u^4)
$, where $u=M/r$ and $b = p_\phi / (-p_t M)$ is the impact parameter.
A circular orbit satisfies $\mathcal{R} = 0 = \mathcal{R}'$; the orbit is stable if $\mathcal{R}'' < 0$.
Values of $b$ leading to repeated roots of $\mathcal{R}$ may be found by solving $\Delta_u(\mathcal{R}) = 0$ for $b$, where $\Delta_u$ denotes the discriminant. Phase boundaries in the $(q^2,a^2)$-plane are found by setting `the discriminant of the discriminant' to zero; remarkably, this expression factorizes as follows:
\beq
\Delta_b \left[ \Delta_u \left(\mathcal{R}\right) / (b-a)^6 \right]  = 2^{32} \left(1 - a^2 - q^2\right) \left(27 a^2 - q^2 (9 - 8q^2)^2  \right)^3.
\eeq

\begin{figure}
\centering
 \includegraphics[height=7cm]{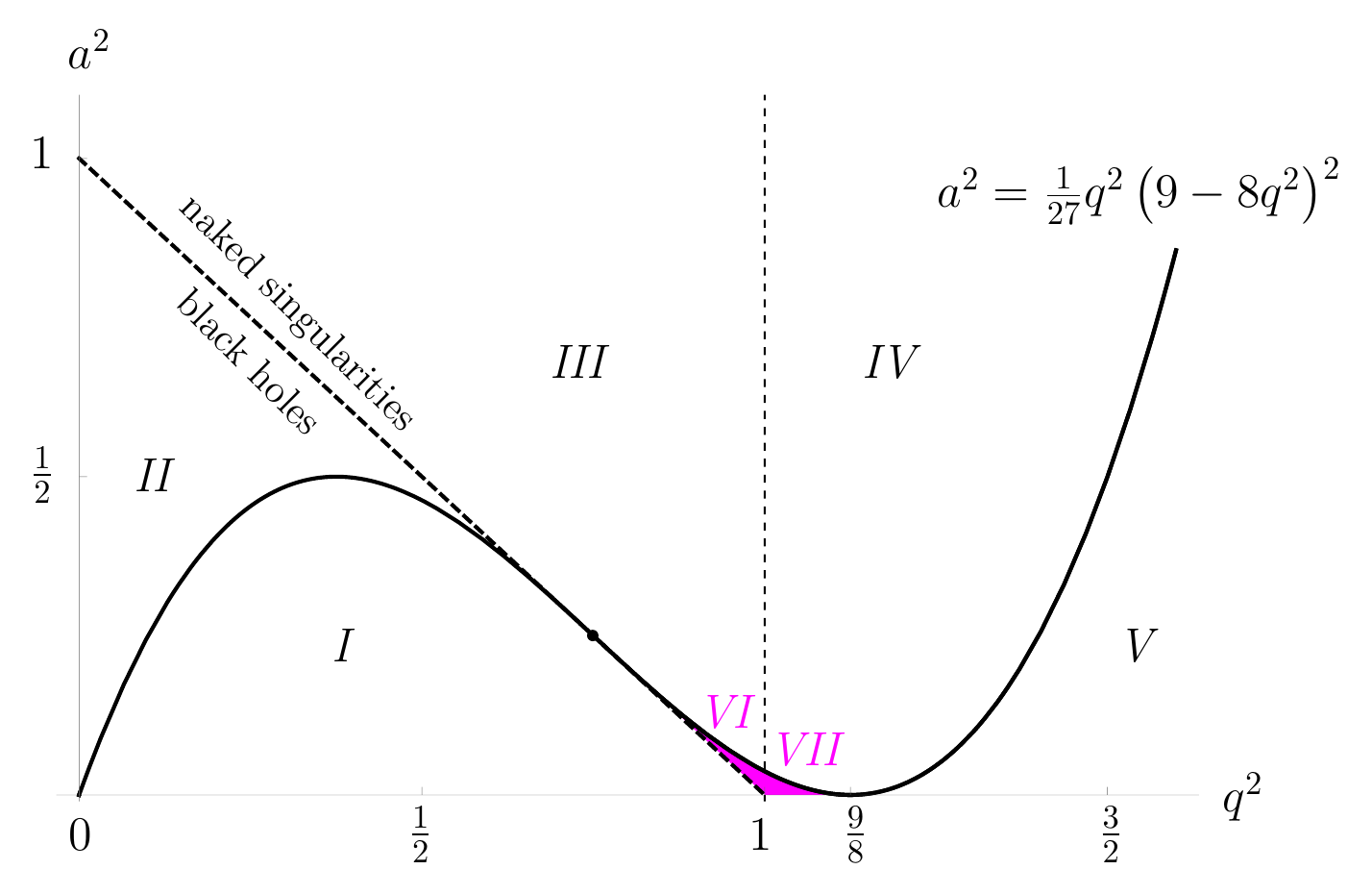}
 \caption{Phase diagram for equatorial circular photon orbits of positive radius ($r>0$) in Kerr--Newman spacetimes (cf.~Fig.~2.1 in \cite{Balek:1989}; also Fig.~6 in \cite{Stuchlik:1981} and Fig.~4 in \cite{Pugliese:2013zma}). {\it All orbits are unstable except where noted to the contrary}. Region $I$: two exterior ($r>r_{h+}$) orbits. $II$: two interior ($0<r<r_{h-}$) and two exterior orbits; the innermost orbit is stable.  $III$ \& $IV$: two orbits; the inner orbit is stable. $V$: no orbits. $VI$ \& $VII$: four orbits; the inner pair are stable. Cases $III$ and $VI$ admit a counter-rotating orbit; $IV$ and $VII$ do not. For $a^2=0$, $0\leq q <1$ (Reissner--Nordstr\"om \& Schwarzschild), one exterior orbit. For $a^2=0$, $1<q^2<9/8$, two orbits, the inner one is stable. For $q^2=0$, $0<a^2<1$ (Kerr) there are two exterior and one interior orbits for $0<a^2<1$.  In the extremal case $a^2+q^2=1$ there are three orbits: for $0<a<1/2$ one stable horizon orbit ($r=M$) and two exterior orbits; for $1/2 < a < 1$ one stable interior orbit ($r<M$), one horizon orbit, and one exterior orbit \cite{Ulbricht:2015vwa, Khoo:2016xqv}. At $a=1/2$, $q=\sqrt{3}/2$, where regions $I$, $II$, $III$ and $VI$ meet, there is a marginally stable ($\mathcal{R}''=0$) horizon orbit and one exterior orbit at $r=3M$.}
 \label{fig:phase-diagram}
\end{figure}

Figure \ref{fig:phase-diagram} shows the Balek--Bi\v{c}\'ak--Stuchl\'ik \cite{Balek:1989} phase diagram for ECPOs in the charge-versus-spin plane. Stable ECPOs exist within the inner horizon $r_{h-}$ in the black hole regime ($a^2+q^2<1$, $r_{h\pm} = M (1 \pm \sqrt{1 - a^2 - q^2})$) in region $II$, on (or inside) extremal horizons for $a<1/2$ ($a>1/2$), and around naked singularities in regions $III$, $IV$ and $VI$.

\subsection{Stable photon orbits of di-holes: existence\label{sec:dihole}}

\subsubsection{Geodesic equations and Hamiltonian formalism}
The exterior of a stationary axisymmetric spacetime in electrovacuum is described in Weyl--Lewis--Papapetrou coordinates \cite{Stephani:2003tm} $\{t,\rho,z,\phi\}$ by the line element
\beq
ds^2 = -f(dt - w d\phi)^2 + f^{-1} \left[ e^{2 \gamma} (d\rho^2 + dz^2) + \rho^2 d\phi^2 \right] , \label{eq:wlp}
\eeq
where $f$, $\gamma$ and $w$ are functions of $\rho$ and $z$ only. Its geodesics $x^{\mu}(\lambda)$ are the integral curves of Hamilton's equations, with $H(x^{\mu},p_{\mu}) = \frac{1}{2} g^{\mu \nu}(x) p_{\mu} p_{\nu}$, where $p_{\mu} \equiv g_{\mu \nu} \frac{dx^\nu}{d\lambda}$, $\lambda$ is an affine parameter, and $H$, $p_\phi$ and $p_t$ are constants of motion. In the null case, $H = 0$, one may set $p_t = -1$ without loss of generality, by availing of the affine-rescaling freedom ($\lambda \rightarrow \alpha \lambda$). Null geodesics are invariant under a conformal transformation of the metric, $g_{\mu \nu} \rightarrow \Omega^2(x) g_{\mu \nu}$, and so one may recast the Hamiltonian in the following canonical 2D form,
\beq
H = \frac{1}{2} \left(p_\rho^2 + p_z^2 \right) + \mathcal{U}, \qquad
\mathcal{U}(\rho,z) = -\frac{1}{2} e^{2\gamma} \left( f^{-2} - \frac{(p_\phi - w)^2}{\rho^2} \right). \label{Hamiltonian}
\eeq
Null orbits of constant $\rho,z$ arise where $\mathcal{U} = 0 = \bnab \mathcal{U}$, with $\bnab \equiv (\frac{\partial}{\partial \rho}, \frac{\partial}{\partial z})$. Such orbits are stable if the stationary point is a local minimum of $\mathcal{U}$, that is, if $\det \mathcal{H}(\mathcal{U}) > 0$ and $\text{Tr} \, \mathcal{H}(\mathcal{U}) > 0$, where $\mathcal{H}(\mathcal{U})$ denotes the Hessian matrix for $\mathcal{U}(\rho,z)$. More generally, photon orbits will be kinematically trapped wherever there exists a closed contour $\mathcal{U}=0$ enclosing a region in which $\mathcal{U}$ is negative on some open set.

The potential can be factorized into $\mathcal{U} = -\frac{e^{2\gamma}}{2\rho^2} (h^+ - p_\phi) (h^- + p_\phi)$, where
\beq
h^{\pm}(\rho, z) \equiv \rho f^{-1} \pm w . \label{eq:hpm}
\eeq
Thus, it is sufficient to seek closed contours $h^+ = p_\phi$ or $h^- = -p_\phi$. As $p_\phi$ may take any positive or negative value, closed contours $\mathcal{U}=0$ -- and thus stable photon orbits -- exist in the vicinity of any local maximum of $h^\pm$. (In the static case, $w = 0$ and we write $h \equiv h^\pm$).

\subsubsection{Majumdar--Papapetrou di-holes}

A Majumdar--Papapetrou (MP) \cite{Majumdar:1947eu,Papapetrou:1947} di-hole comprises a pair of extremally charged Reissner--Nordstr\"{o}m black holes in static equilibrium \cite{Chandrasekhar:1989vk}. The spacetime geometry is described by (\ref{eq:wlp}) with $\gamma = w = 0$ and $f^{-1/2} = 1 + M_+ / r_+ + M_- / r_-$, where $M_\pm = (1 \pm \eta) M_0 / 2$, $r_{\pm} = \sqrt{\rho^2 + (z - z_\pm)^2}$, $z_\pm = \pm M_\mp d / (M_+ + M_-)$, where $d$ is the coordinate separation of the event horizons \cite{Hartle:1972ya}, $\eta$ parameterizes the mass ratio, and $M_{0}$ is the total mass.

The equal-mass MP di-hole system ($\eta=0$, $M \equiv M_\pm$) possesses SPOs for black hole separations in the range $\sqrt{16/27} \le d / M \le \sqrt{32/27}$ \cite{Shipley:2016omi}. In this regime, $h$ possesses four stationary points: a local maximum at $\rho=\rho_{(1)}$, $z=0$, $h=p_\phi^{(1)}$; a saddle at $\rho=\rho_{(2)} > \rho_{(1)}$, $z=0$, $h=p_\phi^{(2)} < p_\phi^{(1)}$, and a pair of saddles above and below the plane at $\rho=\rho_{(3)} < \rho_{(1)}$, $z = \pm z_{(3)}$, $h=p_\phi^{(3)} < p_\phi^{(1)}$. For $d > M$, we have $p_\phi^{(2)} > p_\phi^{(3)}$ and thus rays with $p_\phi$ between $p_\phi^{(2)}$ and $p_\phi^{(3)}$ may connect to infinity but not to the black holes; for $d < M$, the order swaps, $p_\phi^{(3)} > p_\phi^{(2)}$, and so the converse holds.

In the special case $d = M$ we have $p_\phi^{(2)} = p_\phi^{(3)}$ and thus the local maximum at $\rho_{(1)} = \sqrt{3}M/2$, $z=0$ with $p_{\phi}^{(1)} = 9\sqrt{3}M/2$ is enclosed by a single contour, $h = p_{\phi}^{(2)} = p_{\phi}^{(3)} \equiv \frac{1}{2} 5^{5/4} \varphi^{3/2} M$, which connects three saddle points, at $\rho_{(2)} = \frac{1}{2} 5^{1/4} \varphi^{3/2}M$, $z=0$ and $\rho_{(3)} = \frac{1}{2} 5^{1/4} \varphi^{-1/2}M$, $z_{(3)}=\pm M/(2\varphi)$, where $\varphi = \frac{1}{2}(1 + \sqrt{5})$ is the Golden Ratio (see Appendix B in Ref.~\cite{Shipley:2016omi}).

\begin{figure}
 \begin{center}
 \includegraphics[height=7cm]{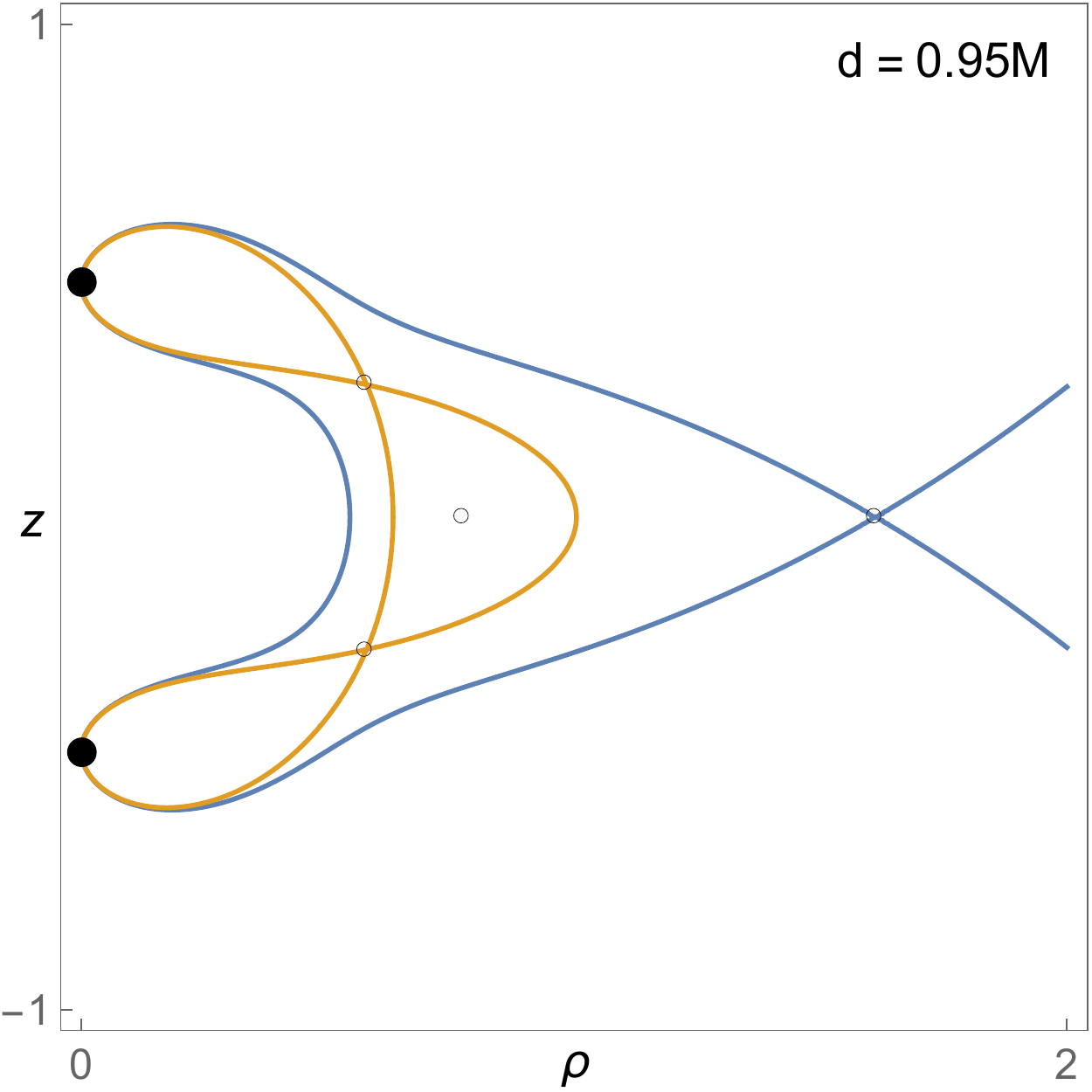} %
 \includegraphics[height=7cm]{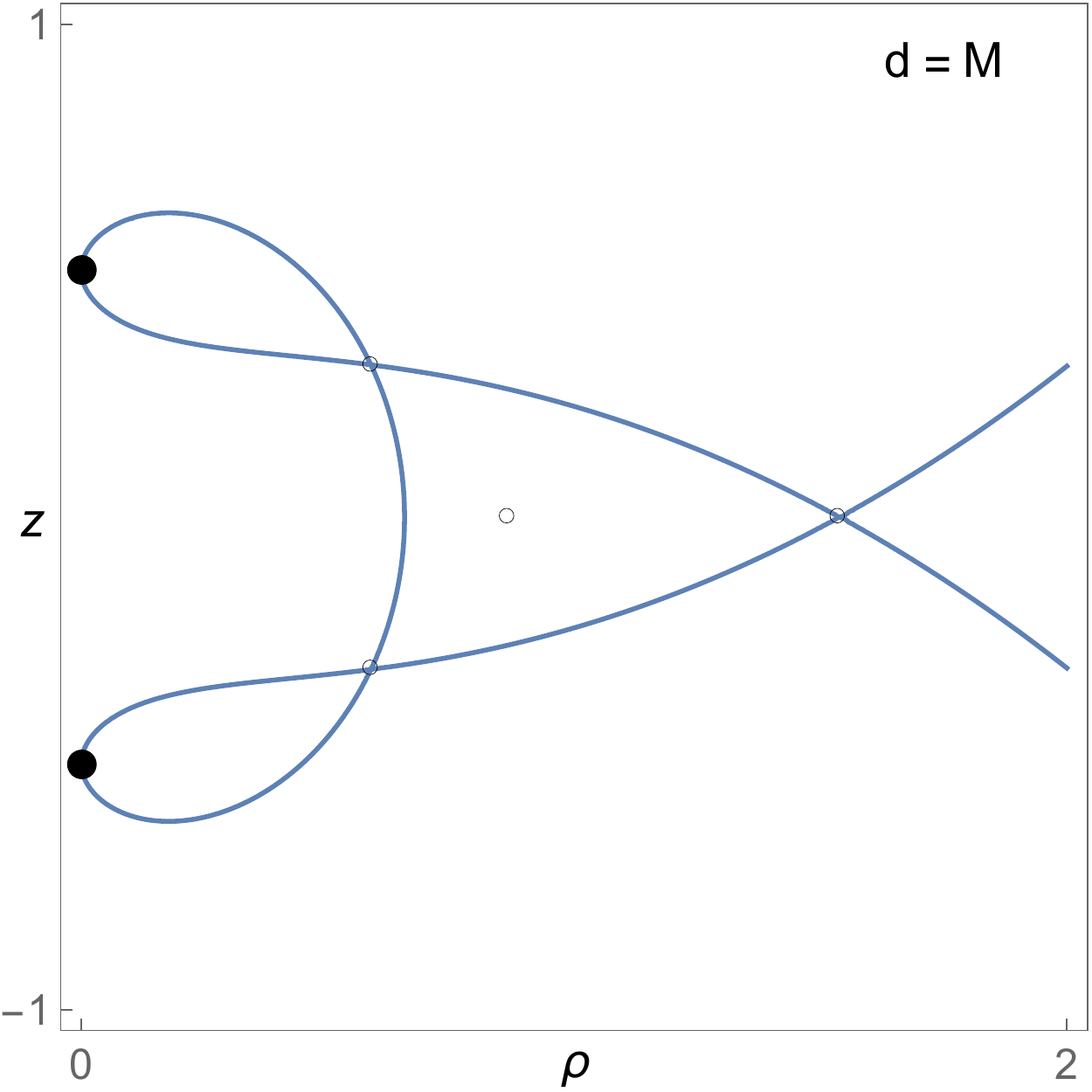} \\
 \includegraphics[height=7cm]{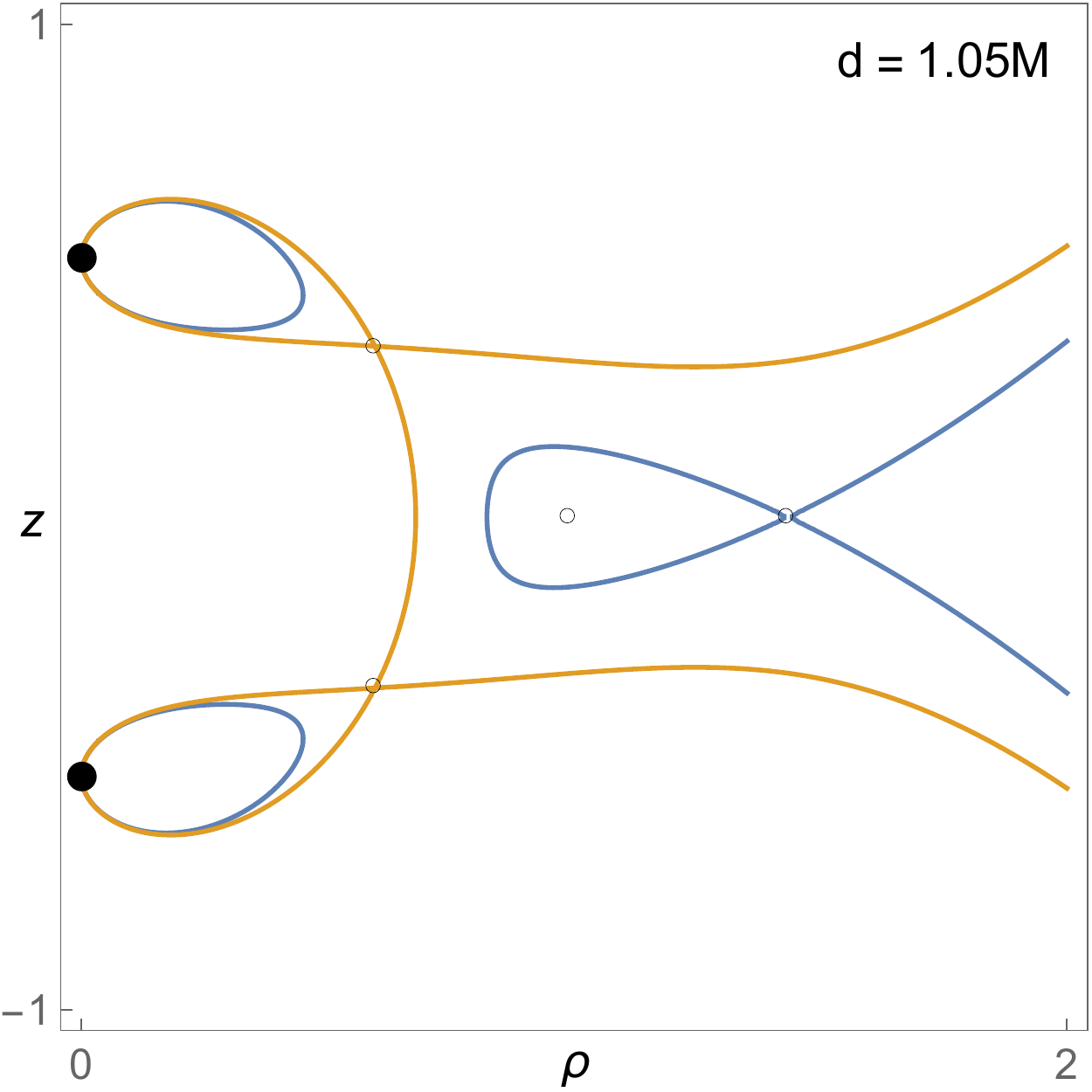}  %
 \includegraphics[height=7cm]{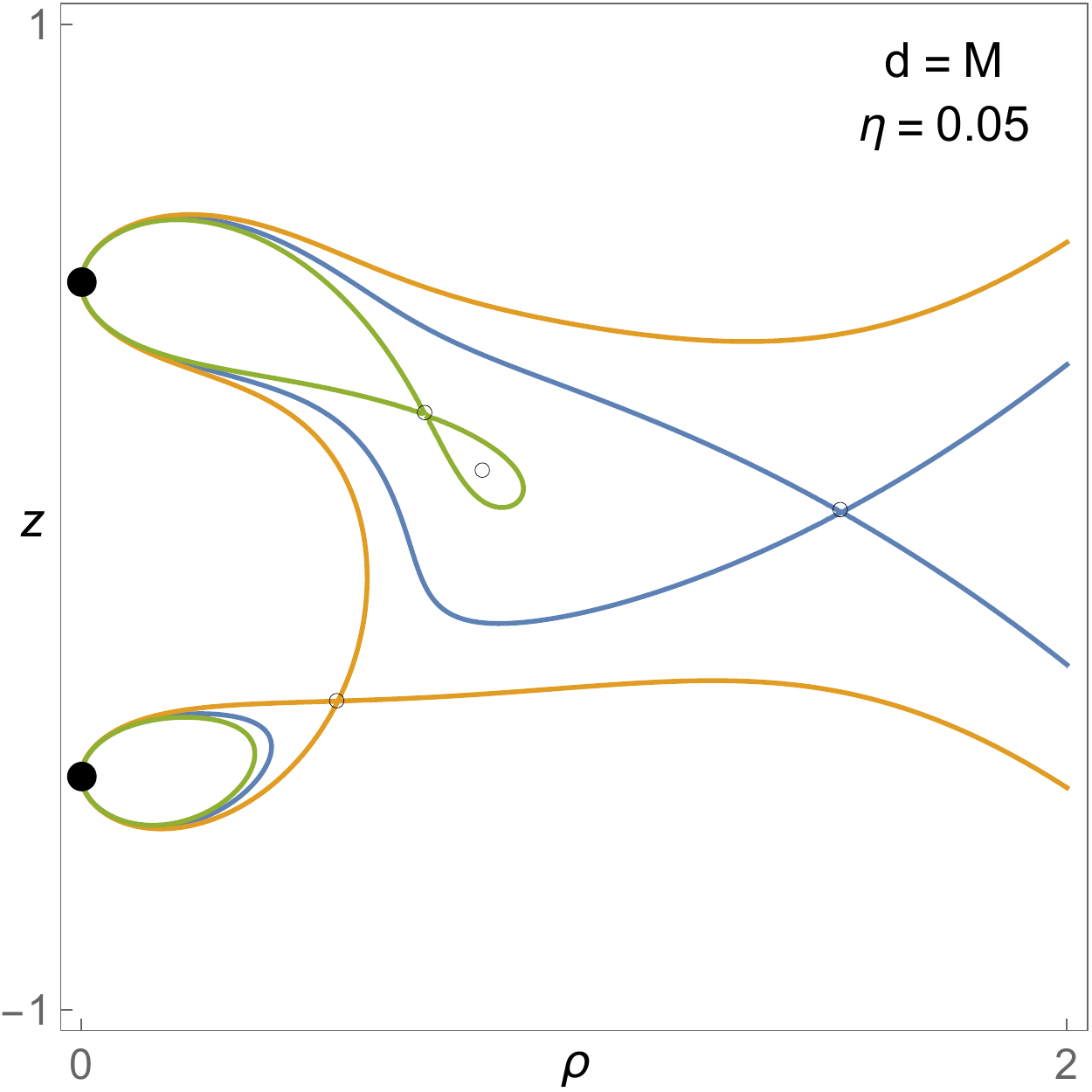}
 \end{center}
 \caption{Contour plots of the height function $h = \rho U^{2}$ for Majumdar--Papapetrou di-holes separated by coordinate distance $d$. Circles indicate stationary points: saddles and one local maximum. Filled circles represent the black hole horizons. The first three plots show equal-mass cases; the lower right plot shows an unequal mass case, $\eta = \frac{M_+ - M_-}{M_+ + M_-} = 0.05$.}
 \label{fig:heightcontours}
\end{figure}

\begin{figure}
 \begin{center}
 \includegraphics[height=7cm]{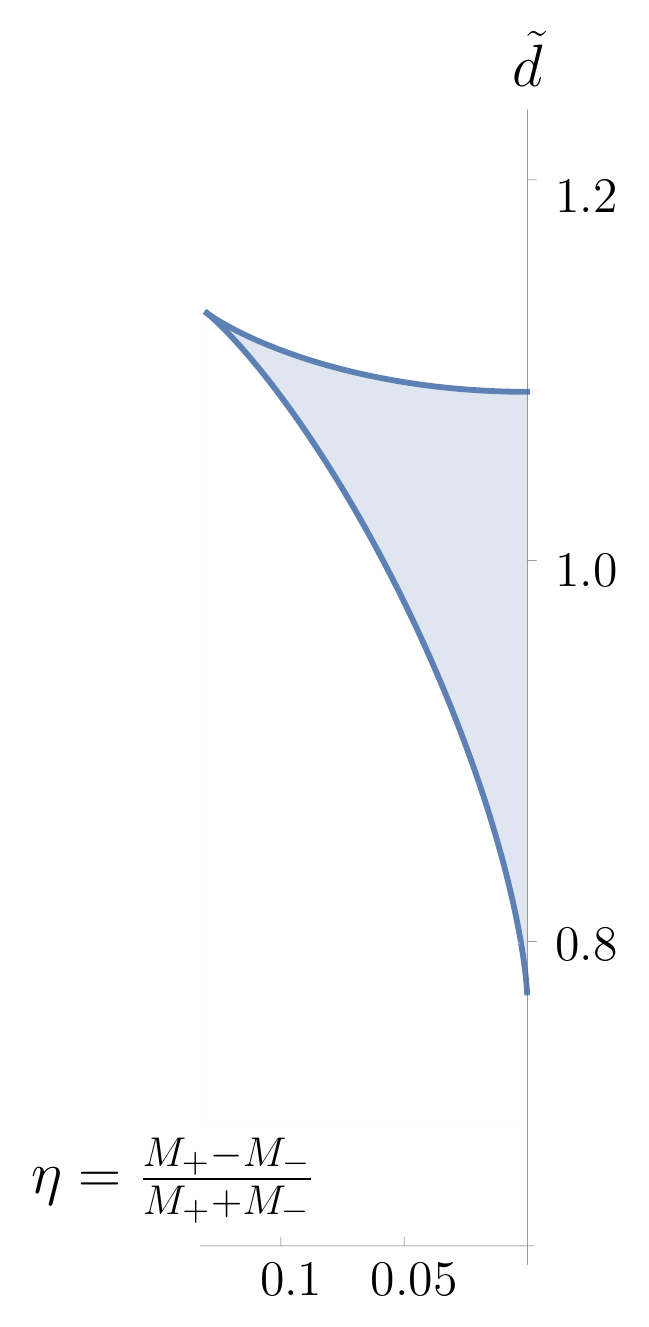}
 \includegraphics[height=7cm]{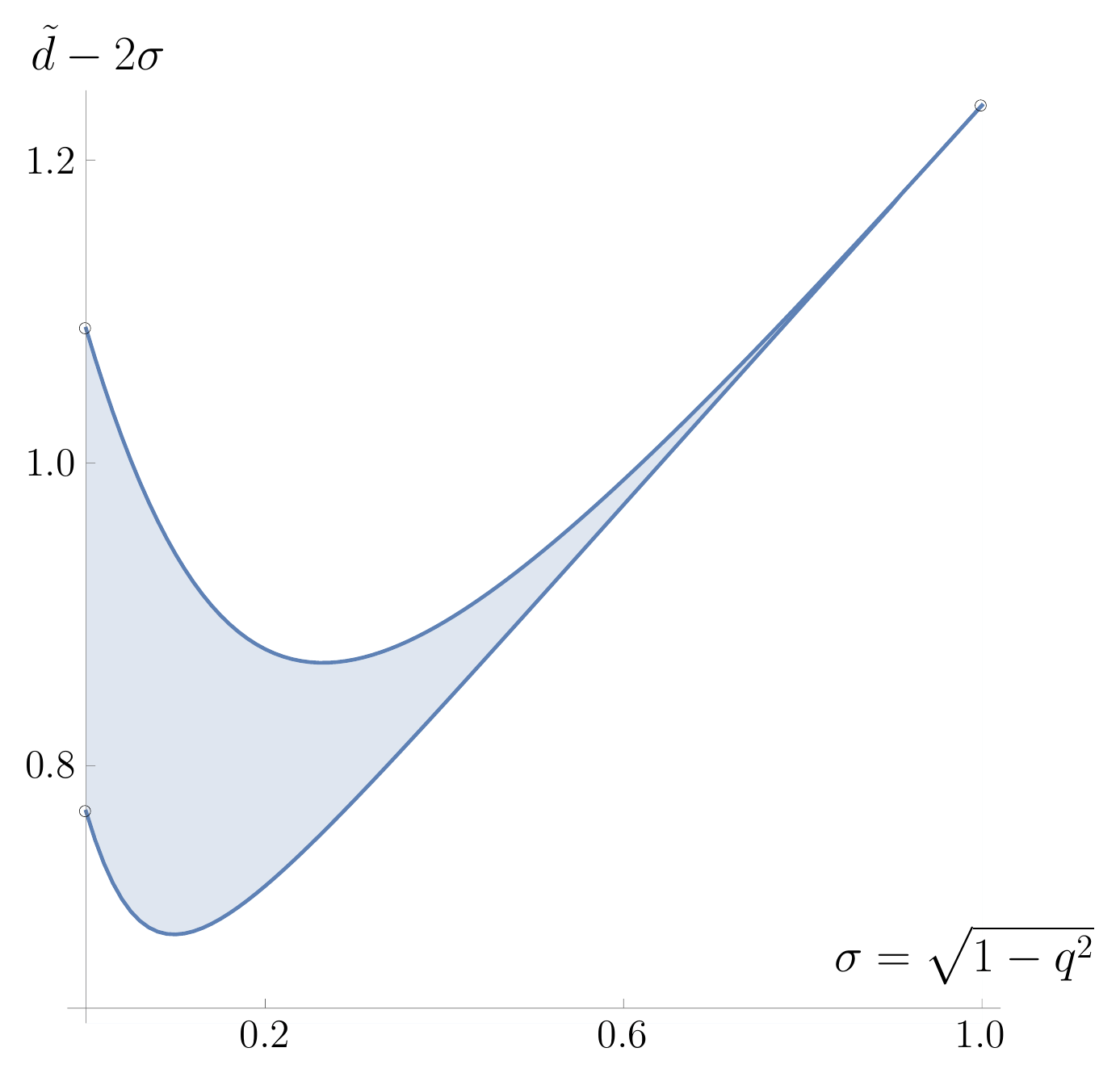}%
 \end{center}
 \caption{Existence regions for stable photon orbits around Reissner--Nordstr\"om di-holes. \\ {\it Left:} Majumdar--Papapetrou di-hole ($Q_\pm = M_\pm$) with mass ratio $\eta$ and (dimensionless) coordinate separation $\tilde{d} = 2 d / (M_+ + M_-)$. {\it Right:} Equal-mass RN black holes with equal charge-to-mass ratios $q = Q_\pm/M_\pm$ held in static equilibrium by a Weyl strut (see text). The vertices shown are at $(0, \sqrt{16/27})$, $(0, \sqrt{32/27})$ \cite{Shipley:2016omi, Wunsch:2013st}, and $(1, 2/\varphi)$ \cite{CoelhoHerdeiro2009}.
 }
 \label{fig:RNdihole}
\end{figure}

Figure \ref{fig:heightcontours} illustrates the kinematically bound regions for SPOs in the $d < M$, $d = M$ and $d > M$ cases. It also shows an unequal-mass case, with four contours, in which the SPO region is dragged towards the more massive partner. Increasing the mass ratio has the effect of diminishing the SPO existence regions, as shown in Fig.~\ref{fig:RNdihole}. Beyond $\eta \gtrsim 0.13$, SPOs do not exist.

\subsubsection{Reissner--Nordstr\"om di-holes}

The $N=2$ Bret\'{o}n--Manko--Aguilar class of electrostatic solutions \cite{Perry:1996ja, BMA:1998, Varzugin:2001ab, Alekseev:Belinski:2007, Manko:2007} is spanned by five parameters: two masses $M_\pm$, two charges $Q_\pm$ and the separation of centres $d$. We focus on the Reissner--Nordstr\"om di-hole sub-class, in which both bodies are black holes ($Q_\pm \le M_\pm$); this sub-class includes the MP di-holes ($M_\pm = Q_\pm$) and Weyl--Bach \cite{Bach:2012} di-holes ($Q_\pm = 0$) as special cases. With the exception of the MP cases, the charged black holes are held in equilibrium by a `Weyl strut' \cite{Bach:2012}, and their horizons appear as `rods' of coordinate length $2 \sqrt{M_\pm^2 - Q_\pm^2}$ on the symmetry axis, in coordinate system (\ref{eq:wlp}).

We examined a two-parameter sub-family: the equal-mass, equal-charge black holes with $q  = Q_\pm / M_\pm$ and $M_\pm = M$, held in equilibrium by a Weyl strut imparting a force $F = G \left[ d^2/(2M\sigma)^2 - 1\right]^{-1}$ \cite{Alekseev:Belinski:2007, Manko:2007} where $\sigma = \sqrt{1-q^2}$. The coordinate distance between the horizons is $d - 2 \sigma M$. Using a numerical root finder, we find that SPOs can exist all the way up to, but not including, the uncharged (Weyl--Bach \cite{Bach:2012}) limit. The existence region for SPOs is shown in Fig.~\ref{fig:RNdihole}, together with analytic values for the MP and Weyl--Bach cases.

\subsection{Stable photon orbits of di-holes: geodesic structure\label{sec:poincare}}

The 2D Hamiltonian system (\ref{Hamiltonian}) is non-integrable, in general, and thus SPOs may possess rich structure.

We introduce a dimensionless parameter $\mu$ via
\beq
p_{\phi}(\mu) = \mu p_{\phi}^{(1)} + (1 - \mu) p_{\phi}^{*}, \label{eq:mu}
\eeq
where $p_\phi^{*}=\text{max}(p_\phi^{(2)}, p_\phi^{(3)})$, noting that kinematically bounded SPOs exist in the range $0 < \mu < 1$. In the limiting regime $\mu \rightarrow 1^-$, the bound region is a small ellipse in the $(\rho, z)$-plane, with $\mathcal{U}$ of Eq.~(\ref{Hamiltonian}) resembling the potential of an anisotropic harmonic oscillator. We note that $\mathcal{U}_{,zz} > \mathcal{U}_{,\rho\rho}$ for $d > M$, and the converse for $d < M$. Generically, SPOs for $\mu \rightarrow 1^-$ will describe precessing ellipses in a small elliptical region of the $(\rho, z)$-plane. However, even in this regime, secular effects from higher-order corrections to $\mathcal{U}$ cannot be neglected if the frequencies are commensurate ($\sqrt{\mathcal{U}_{,\rho\rho} / \mathcal{U}_{,zz}} \approx n_{1} / n_{2}$, where $n_{1}$ and $n_{2}$ are small positive integers). In particular, the `isotropic' case $d = M$ (with $\mathcal{U}_{,\rho\rho} = \mathcal{U}_{,zz}$ in the limit $\mu \rightarrow 1^-$) exhibits a 1:1 resonance, and deserves some special consideration.

\begin{figure}
 \begin{center}
 \includegraphics[height=7cm]{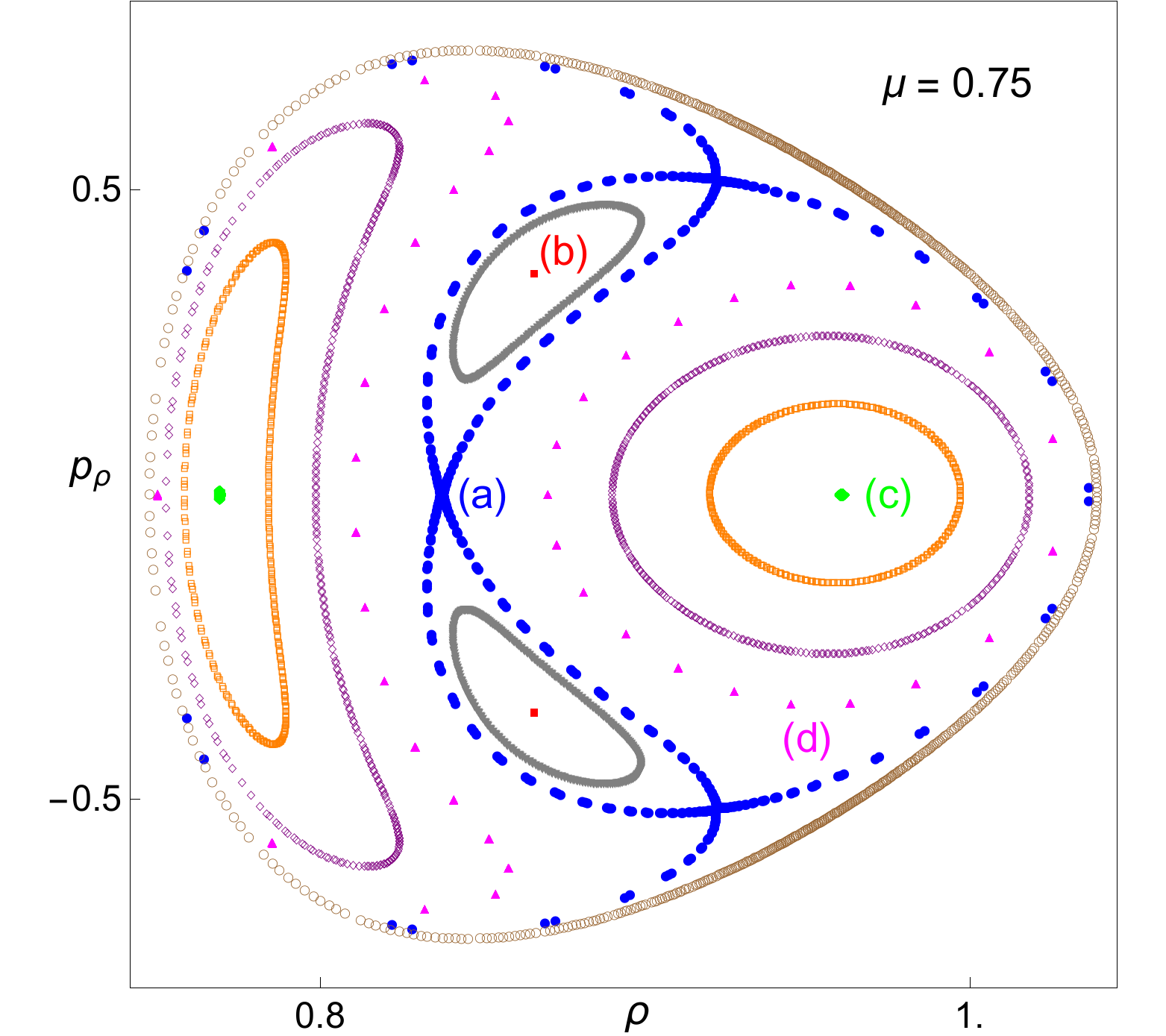}%
 \includegraphics[height=7cm]{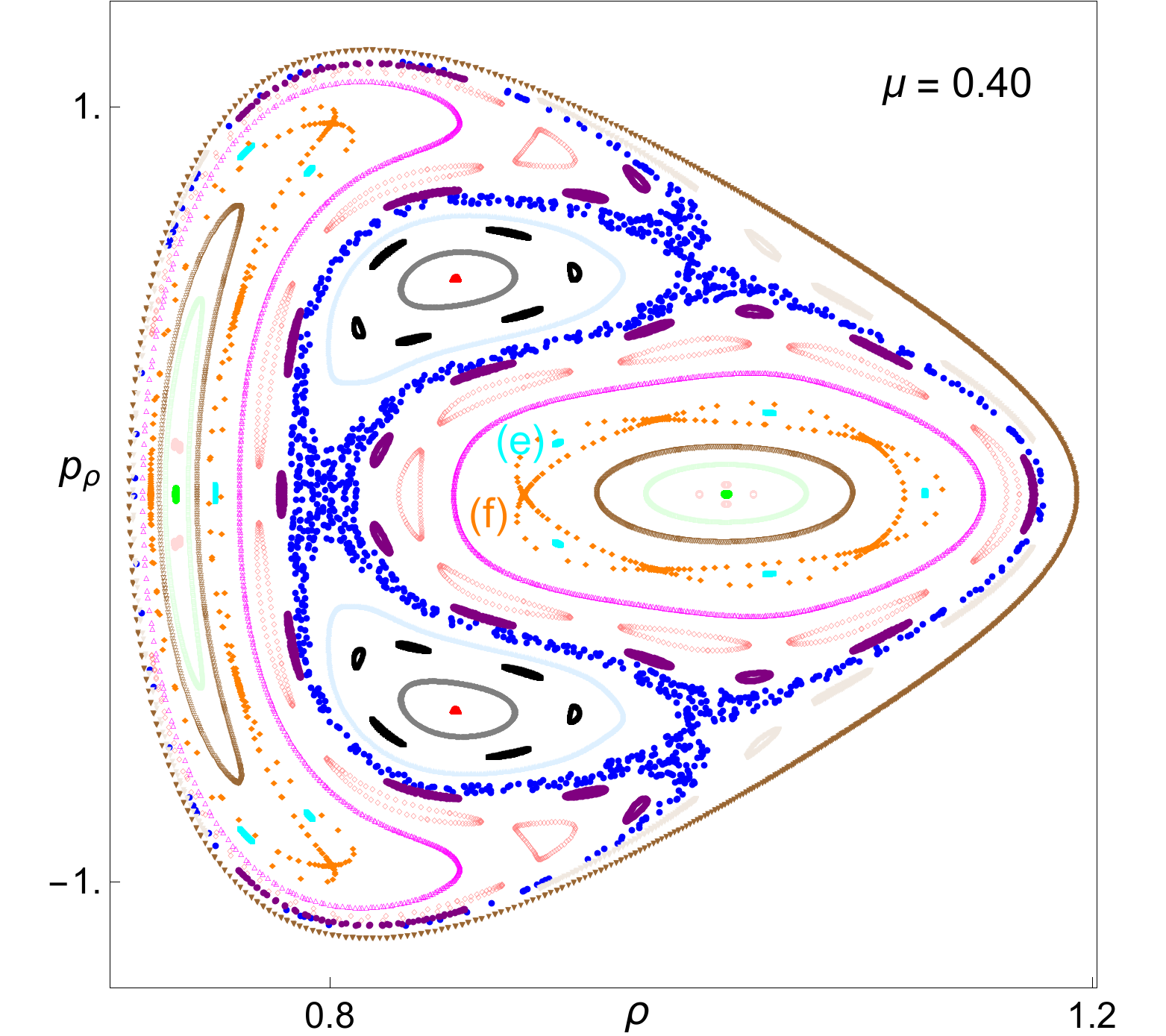} \\
 \includegraphics[height=5cm]{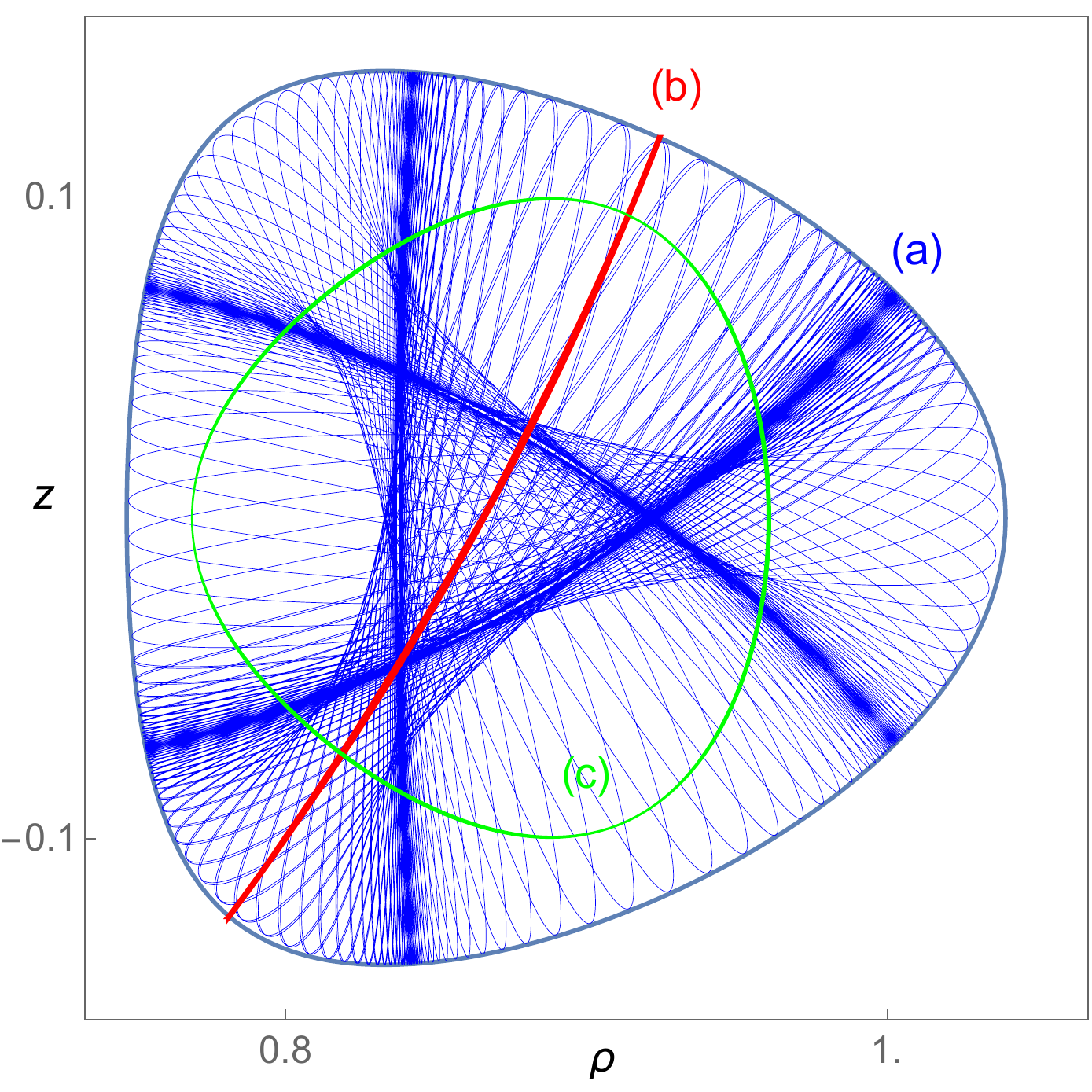}
 \includegraphics[height=5cm]{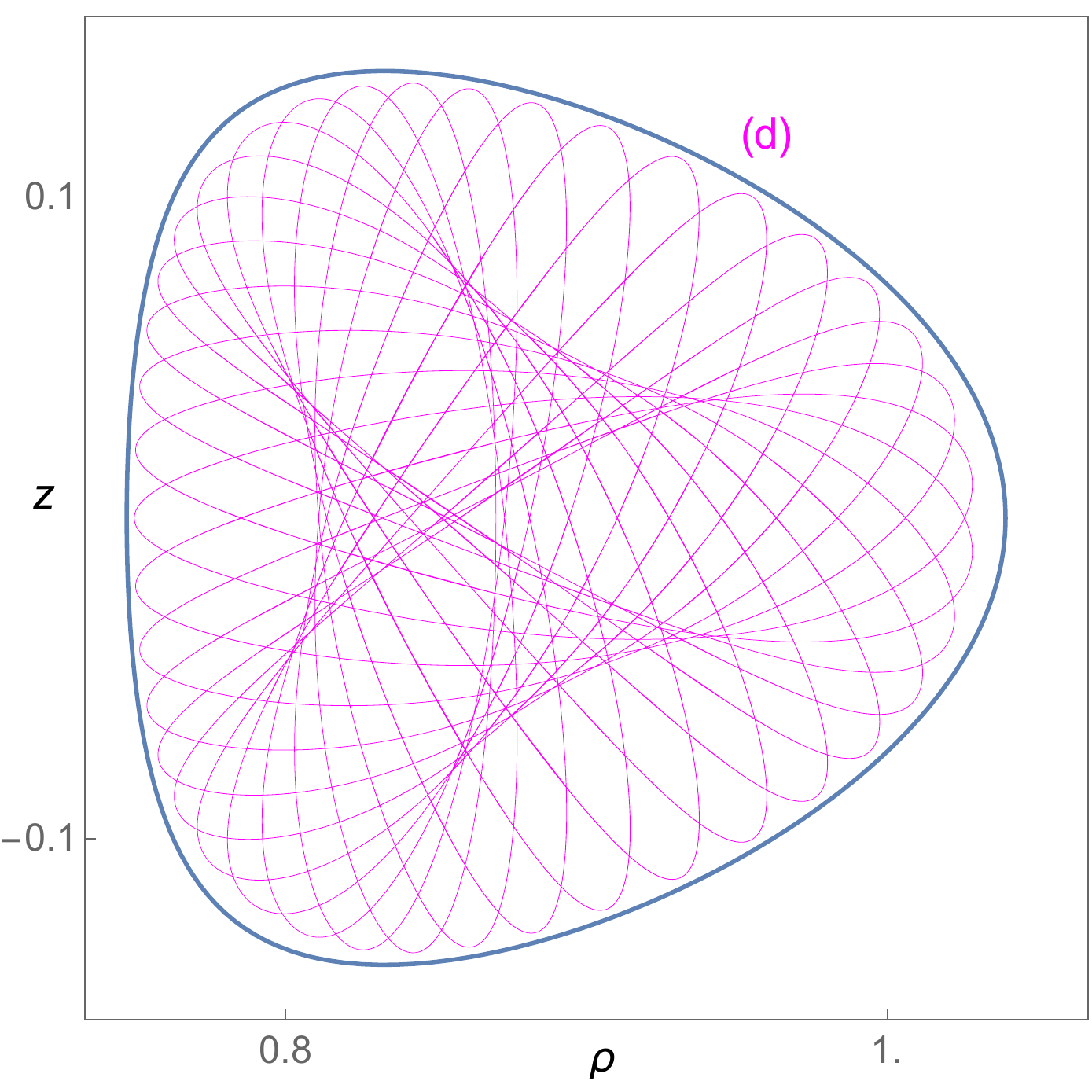}
 \includegraphics[height=5cm]{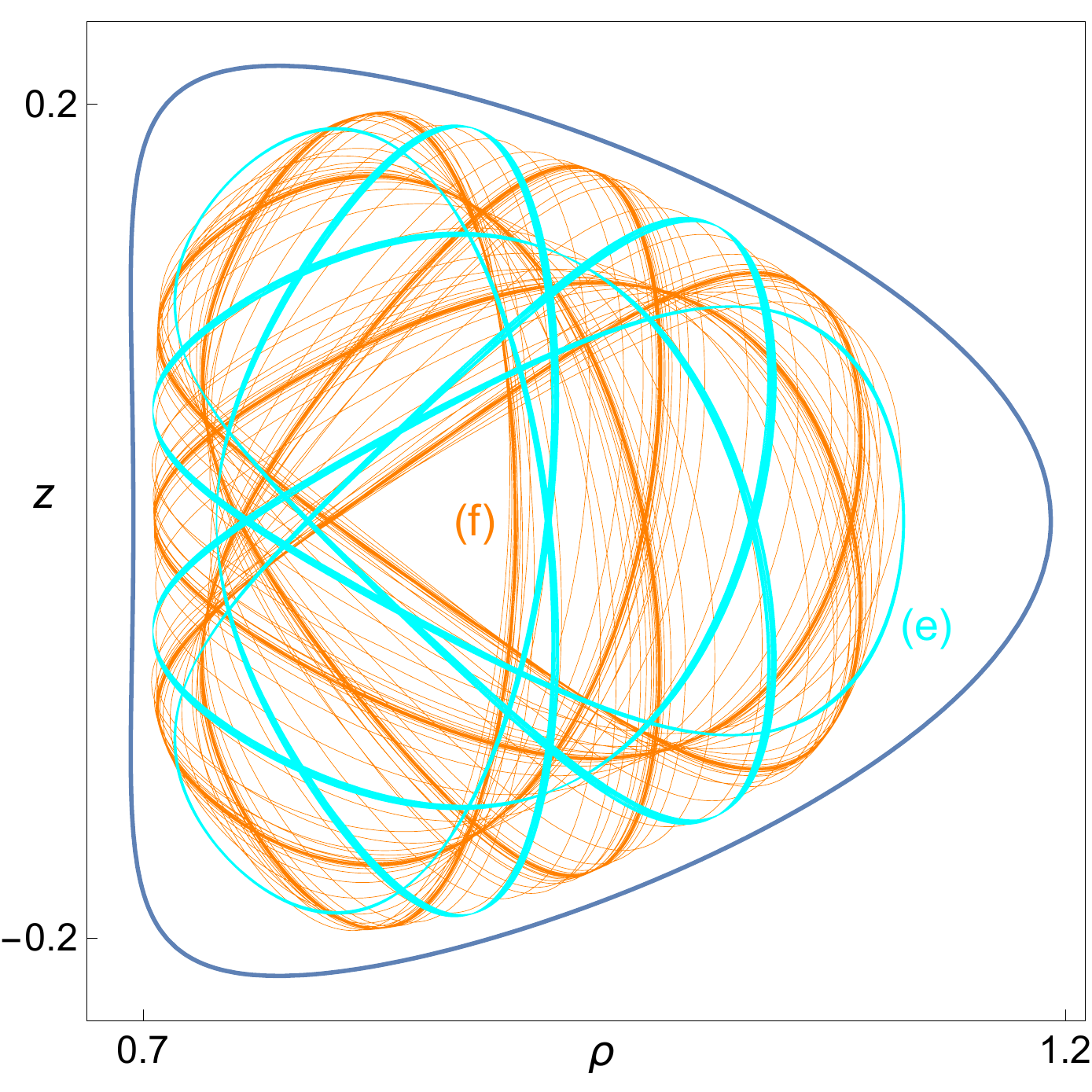} \\
 \includegraphics[height=7cm]{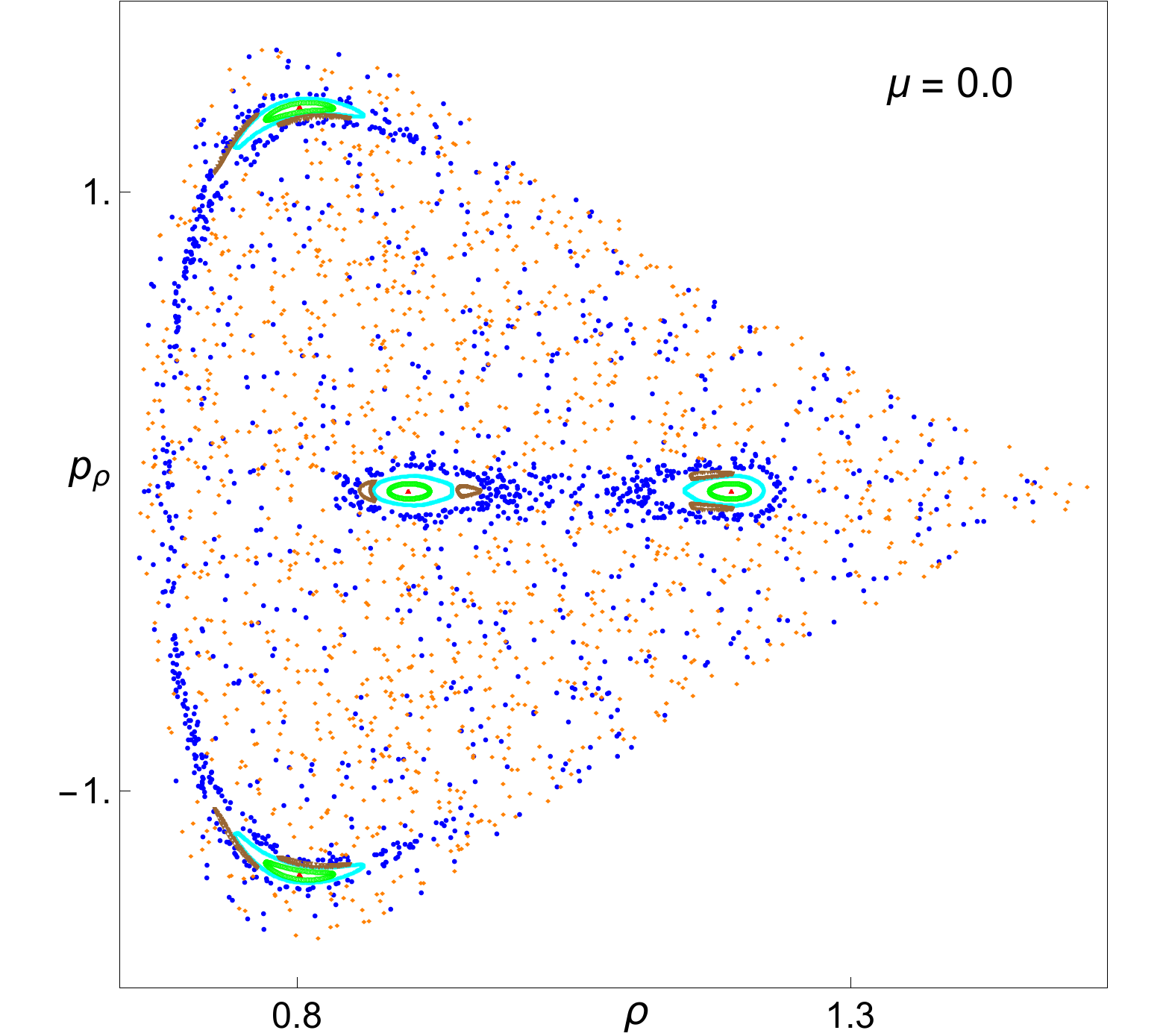} %
 \includegraphics[height=7cm]{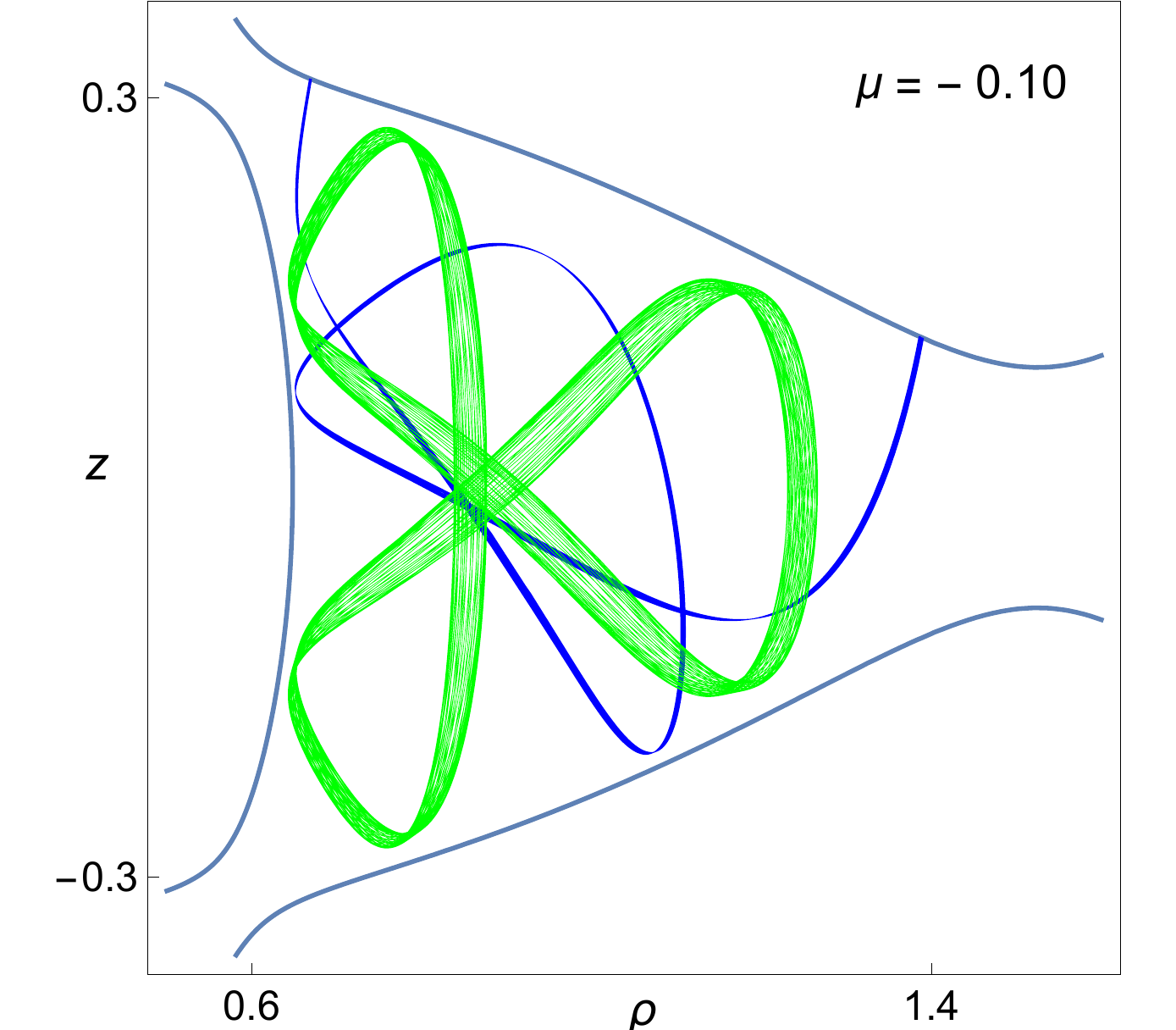}
 \end{center}
 \caption{Poincar\'e sections [taken in the $z=0$ plane] and SPOs [viewed in the $(\rho,z)$-plane with $\phi$ suppressed] for the equal-mass Majumdar--Papapetrou di-hole in the special case $d=M$.}
 \label{fig:poincare}
\end{figure}

Figure \ref{fig:poincare} shows a selection of Poincar\'e sections for SPOs kinematically bounded in a toroidal  region around the $d=M$ MP di-hole. For $\mu = 0.75$, a separatrix (a) connects three saddle points and divides between `libration' (b) and `rotation' (c); in either region there exist high-order resonances (box orbits), such as (d). Chaos is confined to very narrow bands in this section. For $\mu = 0.4$, the separatrix has degenerated into a chaotic band, and stable (e) and unstable (f) KAM tori are manifest around resonances \cite{Berry:1978}. As $\mu$ is decreased, further lower-order resonances branch off from 1:1 resonances [(b) and (c)], and the chaotic regions grow. For $\mu = 0$, chaos is dominant. Interestingly, SPOs can persist even into the kinematically unbound regime, $\mu < 0$; Fig.~\ref{fig:poincare} shows examples of libration- and rotation-type SPOs for $\mu =  -0.1$.

Remarkably, the qualitative features seen in Fig.~\ref{fig:poincare} are shared by the well-studied H\'enon--Heiles (HH) system \cite{Henon:1964, Henon:1983, Berry:1978}: a 2D non-integrable dynamical system with a cubic potential $\mathcal{U}_{{\rm HH}} = \tfrac{1}{2} \left( x^2 + y^2 \right) + x y^2 - \tfrac{1}{3} x^3$ and energy $\mathcal{E}$. Naively, for $\mu \rightarrow 1^-$, one may expand our $\mathcal{U}$ in a power series around its minimum $\mathcal{U}_0 = -\mathcal{E}$, and attempt to map onto a generalized HH system. However, in our naive approach, we found that it was necessary to expand to \emph{quartic} order if we wish to observe the key features in the Poincar\'e section, such as the separatrix (a). This suggests that the link between SPOs of the $d=1$ MP di-hole and orbits of the HH system is rather subtle, and it may repay further investigation.

\begin{figure}
 \begin{center}
 \includegraphics[height=7.25cm]{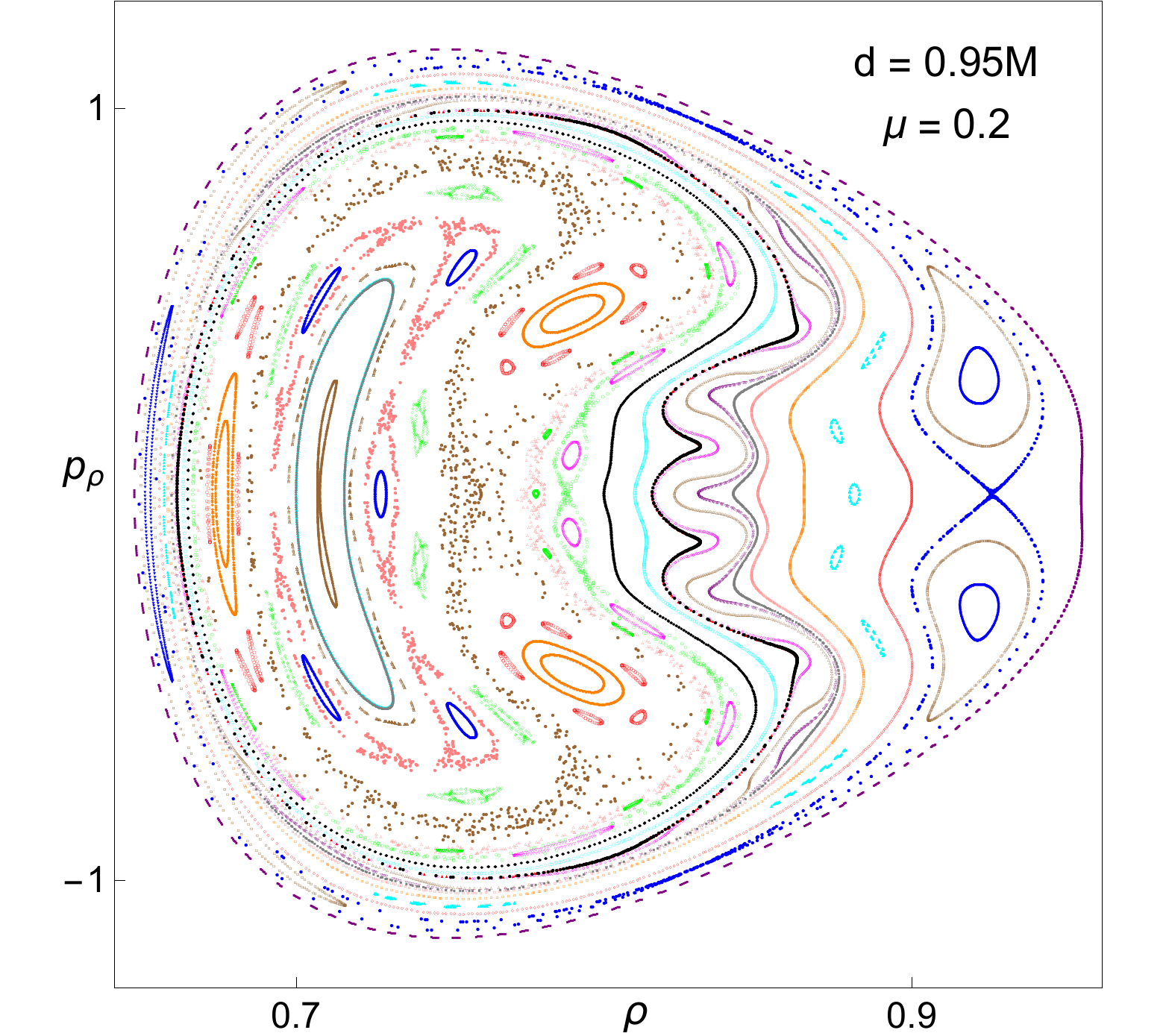} %
 \includegraphics[height=7.25cm]{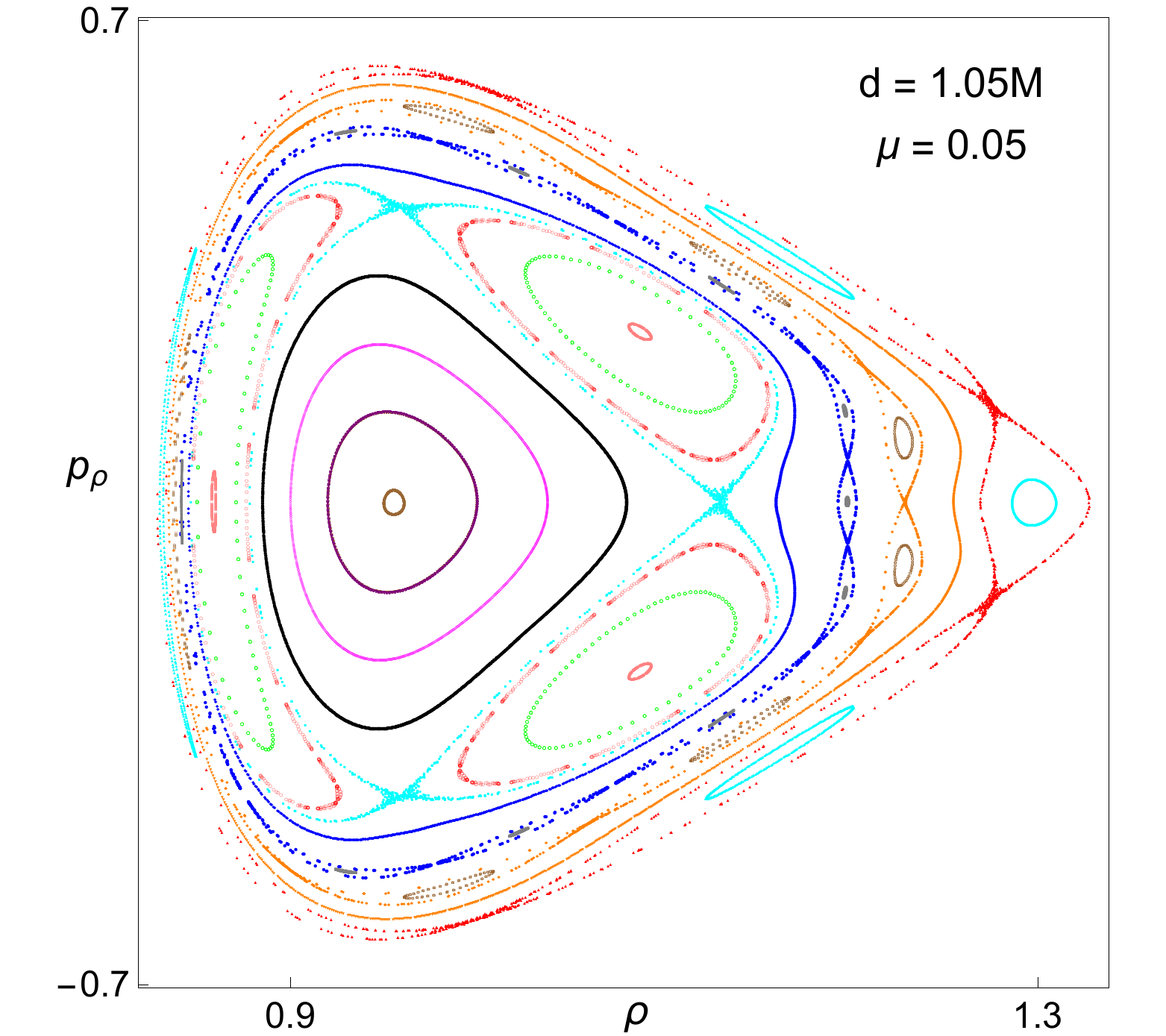}
 \end{center}
 \caption{Poincar\'{e} sections in $z=0$ plane for SPOs on MP di-hole spacetimes with coordinate separations $d = 0.95 M$ (left) and $d = 1.05 M$ (right). See also Fig.~\ref{fig:heightcontours}.}
 \label{fig:aneq1}
\end{figure}

Figure \ref{fig:aneq1} shows Poincar\'e sections for SPOs around MP di-holes with separations less than (left) and greater than (right) $d = M$. These sections exhibit a variety of structure. In addition to libration and vibrational motions, various low-order resonances, and KAM islands embedded within chaotic bands, we see novel features, such as the `crenulations' in the $d=0.95M$ case. As $\mu$ is increased towards $1$, various resonances disappear and the richness of structure declines; as $\mu$ is decreased towards $0$, the chaotic bands become more dominant.

\subsection{Stable photon orbits in stationary axisymmetric electrovacuum\label{sec:existence}}
Now consider a general stationary axisymmetric geometry described by (\ref{eq:wlp}). The bounded regions for null geodesics may be determined from $h^{\pm}(\rho,z)$, defined in Eq.~(\ref{eq:hpm}). As shown in Appendix \ref{app1}, one may employ the electrovacuum field equations to classify the stationary points of $h^\pm$. Our key result is the following: at a stationary point where either $\bnab h^+ = 0$ or $\bnab h^- = 0$, the second derivatives of the corresponding function $h^\pm$ satisfy
\beq
h^{\pm}_{,\rho \rho} + h^{\pm}_{, z z} = -\frac{2}{\rho} W_\pm^2, \qquad \mathbf{W}_\pm \equiv  h^{\pm} \bnab A_t \mp \bnab A_\phi ,  \label{eq:hess}
\eeq
where $A_t$ and $A_\phi$ are components of the electromagnetic four-potential, and $W_\pm^2 = \mathbf{W}_\pm \cdot \mathbf{W}_\pm$. In the vacuum case ($\bnab A_{t} = 0 = \bnab A_{\phi}$), the right-hand side of Eq.~(\ref{eq:hess}) is zero. It follows that, since $\det \mathcal{H}(h^{\pm}) = h^\pm_{,\rho \rho} h^\pm_{,zz} - (h^\pm_{, \rho z})^2 \le 0$, $h^\pm$ cannot possess a (first-order) local maximum, and thus generic SPOs are ruled out. By contrast, for $\bnab A_\mu \neq 0$, the right-hand side is negative, and thus SPOs are possible if $h^\pm_{,\rho\rho}$ is negative and $(h^{\pm}_{, \rho z})^2 < |h^\pm_{,\rho\rho}| \left( 2 W_\pm^2 / \rho - |h^\pm_{,\rho\rho}| \right)$. Where there exists an equatorial symmetry ($h^\pm_{,\rho z} = 0$) the condition for stability reduces to $0 < -h_{,\rho\rho}^\pm < 2 W_\pm^2 / \rho$. In the equal-mass MP case, the stability condition implies that SPOs exist in the range $\sqrt{32/27} > d / M > \sqrt{16/27}$.

\section{Discussion\label{sec:discussion}}
We have shown that, in stationary axisymmetric geometries, (generic) SPOs are forbidden in pure-vacuum environments; but that SPOs may arise in electrovacuum, thanks to the right-hand side of Eq.~(\ref{eq:hess}). Furthermore, our examples (Figs.~\ref{fig:phase-diagram}, \ref{fig:heightcontours}, \ref{fig:RNdihole}) imply that SPOs can exist even in the near-vacuum case ($q \rightarrow 0$), provided that other parameters are appropriately tuned. Figures \ref{fig:poincare} and \ref{fig:aneq1} reveal that SPOs may possess rich geodesic structure even in simple models.

The absence of SPOs for (static, axisymmetric) \emph{vacuum} geometries was noted by Liang in 1974 \cite{Liang:1974}, who remarked of equatorial photon orbits that ``stability within the plane implies instability off the plane'' \footnote{This may also be inferred by taking the limit $v\rightarrow1$ of Eq.~(32) in Ref.~\cite{Bardeen:1970}}. Equation (\ref{eq:hess}) shows how Liang's implication breaks down when an electromagnetic field is introduced. Further generalization of this argument -- to include other fields, matter sources, or a cosmological constant; to spacetimes with less symmetry; or even to modified theories -- would surely deepen our understanding of one of the key properties of Einstein's theory.

Various heuristic arguments suggest that, generically, spacetimes possessing SPOs are dynamically unstable \cite{Friedman:1978, Pani:2010jz, Cardoso:2014sna, Keir:2014oka, Khoo:2016xqv}. If not, they should be associated with rich phenomenology (see Sec.~\ref{sec:intro}). For example, highly relativistic timelike orbits, expected in the vicinity of SPOs, would provide a \emph{stable} mechanism for Gravitational Synchrotron Radiation \cite{Misner:1972jf, Breuer:1973, Chrzanowski:1974nr}: strongly-beamed, high-multipole, high-frequency gravitational waves. By historical curiosity, the Gravitational Synchrotron possibility was investigated in 1972 \cite{Misner:1972kx, Davis:1972dm, Chitre:1972fv}, the year in which J.~Weber's `Lunar Surface Gravimeter' experiment was included aboard the Apollo 17 mission. Following aLIGO's epoch-making results, the community now has renewed hopes for space-based detectors and serendipitous discoveries in the years ahead.

\acknowledgments
With thanks to Z.~Stuchl\'ik and C.~Herdeiro for enlightening correspondence. S.D.~acknowledges financial support under EPSRC Grant No.~EP/M025802/1, and from the Lancaster-Manchester-Sheffield Consortium for Fundamental Physics under STFC Grant No.~ST/L000520/1. J.S.~acknowledges financial support from the University of Sheffield Harry Worthington Scholarship.

\appendix

\section{Electrovacuum field equations and the stationary point classification\label{app1}}

Below we derive Eq.~(\ref{eq:hess}), the key result of Sec.~\ref{sec:existence}.

In Ref.~\cite{Ernst:1967by}, Ernst presents a formulation of the field equations in the case of a stationary axisymmetric electrovacuum spacetime with line element Eq.~\eqref{eq:wlp}. For our purpose, the relevant Einstein--Maxwell equations are (see Eqs.~(7), (6) and (4) in \cite{Ernst:1967by}):
\begin{subequations}
\begin{eqnarray}
(\bnab f)^{2} - \rho^{-2} f^{4} (\bnab w)^{2} + 2 f (\bnab A_{t})^{2} + 2 \rho^{-2} f^{3} (\bnab A_{\phi} - w \bnab A_{t})^{2} &=& f \nabla^{2}f,   \label{eqn:einmax1} \\
\bnab \cdot \left[ \rho^{-2} f^{2} \bnab w - 4 \rho^{-2} f A_{t} \left( \bnab A_{\phi} - w \bnab A_{t} \right) \right] &=& 0,  \label{eqn:einmax2} \\
\bnab \cdot \left[ \rho^{-2} f \left( \bnab A_{\phi} - w \bnab A_{t} \right) \right] &=& 0. \label{eqn:einmax3}
\end{eqnarray}
\end{subequations}
Here the divergence operator is defined to be $\bnab \cdot \mathbf{F} \equiv \frac{1}{\rho}\partial_{\rho} \left(\rho F_{\rho}\right) + \partial_{z} F_{z}$, where $\mathbf{F} = (F_{\rho}, F_{z})$ is any arbitrary vector field.

We will consider the height functions $h^{\pm}$ for the effective potential, defined in Eq.~(\ref{eq:hpm}).
Taking second derivatives, it is quick to establish that
\beq \label{eqn:hderivatives}
h^{\pm}_{,\rho\rho} + h^{\pm}_{,zz} = - \rho f^{-3} ( f \nabla^{2}f ) - f^{-2} f_{,\rho} + 2 \rho f^{-3} ( \bnab f )^{2} \pm \left( w_{,\rho\rho} + w_{,zz} \right).
\eeq
Expanding and rearranging Eq.~\eqref{eqn:einmax2} we obtain
\beq \label{eqn:wderivatives}
w_{,\rho\rho} + w_{,zz} = \rho^{-1} w_{,\rho} - 2 f^{-1} \bnab f \cdot \bnab w + 4 \rho^{2} f^{-2} \bnab \cdot \left[ \rho^{-2} f A_{t} \left( \bnab A_{\phi} - w \bnab A_{t} \right) \right].
\eeq
Substituting Eq.~\eqref{eqn:einmax1} and \eqref{eqn:wderivatives} into the right-hand side of Eq.~\eqref{eqn:hderivatives} gives
\begin{eqnarray}
h^{\pm}_{,\rho\rho} + h^{\pm}_{,zz} &=& \rho f^{-3} (\bnab f)^{2} \mp 2 f^{-1} \bnab f \cdot \bnab w + \rho^{-1} f (\bnab w)^{2} - f^{-2} f_{,\rho} \pm \rho^{-1} w_{,\rho} - 2 \rho f^{-2} (\bnab A_{t})^{2}  \nonumber \\
 && - 2 \rho^{-1} (\bnab A_{\phi} - w \bnab A_{t})^{2} \pm 4 \rho^{2} f^{-2} \bnab \cdot \left[ \rho^{-2} f A_{t} \left( \bnab A_{\phi} - w \bnab A_{t} \right) \right].  \label{eqn:hderivatives2}
\end{eqnarray}

Now, let us assume that we have found a stationary point of either $h^+$ or $h^-$. The stationary point conditions $h^{\pm}_{,\rho} = 0 = h^{\pm}_{,z}$ imply that
\beq \label{eqn:stationarypointconditions}
w_{,\rho} = \pm \left( \rho f^{-2} f_{,\rho} - f^{-1} \right), \qquad w_{,z} = \pm \rho f^{-2} f_{,z}  .
\eeq
In either case ($\pm$), it is possible to eliminate terms involving derivatives of $w$ from Eq.~\eqref{eqn:hderivatives2}. In particular, for stationary points of $h^{\pm}$ we have
\beq \label{eqn:derivw1}
2 f^{-1} \bnab f \cdot \bnab w = \pm \left( 2 \rho f^{-3} (\bnab f)^{2} - 2 f^{-2} f_{,\rho} \right),
\eeq
\beq \label{eqn:derivw2}
\rho^{-1} f (\bnab w)^{2} = \rho f^{-3} (\bnab f)^{2} - 2 f^{-2} f_{,\rho} + \rho^{-1} f^{-1}.
\eeq
Using Eq.~\eqref{eqn:derivw1}, Eq.~\eqref{eqn:derivw2} and the expression for $w_{,\rho}$ given in Eq.~\eqref{eqn:stationarypointconditions}, we see that the first five terms on the right-hand side of Eq.~\eqref{eqn:hderivatives2} vanish. Thus, we are left with
\beq \label{eqn:hderivatives3}
h^{\pm}_{,\rho\rho} + h^{\pm}_{,zz} = - 2 \rho f^{-2} (\bnab A_{t})^{2} - 2 \rho^{-1} (\bnab A_{\phi} - w \bnab A_{t})^{2} \pm 4 \rho^{2} f^{-2} \bnab \cdot \left[ \rho^{-2} f A_{t} \left( \bnab A_{\phi} - w \bnab A_{t} \right) \right].
\eeq
Now consider the final term on the right-hand side of Eq.~\eqref{eqn:hderivatives3}. Expanding this term, using the product rule for the divergence operator, yields
\beq \label{eqn:expandderivative}
\begin{split}
4 \rho^{2} f^{-2} \bnab \cdot \left[ \rho^{-2} f A_{t} \left( \bnab A_{\phi} - w \bnab A_{t} \right) \right]
= 4& f^{-1} \bnab A_{t} \cdot \left( \bnab A_{\phi} - w \bnab A_{t} \right) \\
&+ 4 \rho^{2} f^{-2} A_{t} \bnab \cdot \left[ \rho^{-2} f \left( \bnab A_{\phi} - w \bnab A_{t} \right) \right].
\end{split}
\eeq
Now employing the remaining field equation, Eq.~\eqref{eqn:einmax3}, we see that the final term on the right-hand side of Eq.~\eqref{eqn:expandderivative} vanishes. Thus,
\beq \label{eqn:a10}
4 \rho^{2} \bnab \cdot \left[ \rho^{-2} f A_{t} \left( \bnab A_{\phi} - w \bnab A_{t} \right) \right]
= 4 f^{-1} \bnab A_{t} \cdot \left( \bnab A_{\phi} - w \bnab A_{t} \right).
\eeq
Inserting Eq.~(\ref{eqn:a10}) into (\ref{eqn:hderivatives3}) leads to a right-hand side that can be factorized,  yielding the key result of Eq.~(\ref{eq:hess}).

\bibliographystyle{apsrev4-1}

\bibliography{refs}

\begin{thebibliography}{66}%
\makeatletter
\providecommand \@ifxundefined [1]{%
 \@ifx{#1\undefined}
}%
\providecommand \@ifnum [1]{%
 \ifnum #1\expandafter \@firstoftwo
 \else \expandafter \@secondoftwo
 \fi
}%
\providecommand \@ifx [1]{%
 \ifx #1\expandafter \@firstoftwo
 \else \expandafter \@secondoftwo
 \fi
}%
\providecommand \natexlab [1]{#1}%
\providecommand \enquote  [1]{``#1''}%
\providecommand \bibnamefont  [1]{#1}%
\providecommand \bibfnamefont [1]{#1}%
\providecommand \citenamefont [1]{#1}%
\providecommand \href@noop [0]{\@secondoftwo}%
\providecommand \href [0]{\begingroup \@sanitize@url \@href}%
\providecommand \@href[1]{\@@startlink{#1}\@@href}%
\providecommand \@@href[1]{\endgroup#1\@@endlink}%
\providecommand \@sanitize@url [0]{\catcode `\\12\catcode `\$12\catcode
  `\&12\catcode `\#12\catcode `\^12\catcode `\_12\catcode `\%12\relax}%
\providecommand \@@startlink[1]{}%
\providecommand \@@endlink[0]{}%
\providecommand \url  [0]{\begingroup\@sanitize@url \@url }%
\providecommand \@url [1]{\endgroup\@href {#1}{\urlprefix }}%
\providecommand \urlprefix  [0]{URL }%
\providecommand \Eprint [0]{\href }%
\providecommand \doibase [0]{http://dx.doi.org/}%
\providecommand \selectlanguage [0]{\@gobble}%
\providecommand \bibinfo  [0]{\@secondoftwo}%
\providecommand \bibfield  [0]{\@secondoftwo}%
\providecommand \translation [1]{[#1]}%
\providecommand \BibitemOpen [0]{}%
\providecommand \bibitemStop [0]{}%
\providecommand \bibitemNoStop [0]{.\EOS\space}%
\providecommand \EOS [0]{\spacefactor3000\relax}%
\providecommand \BibitemShut  [1]{\csname bibitem#1\endcsname}%
\let\auto@bib@innerbib\@empty
\bibitem [{\citenamefont {Abbott}\ \emph
  {et~al.}(2016{\natexlab{a}})\citenamefont {Abbott} \emph
  {et~al.}}]{Abbott:2016blz}%
  \BibitemOpen
  \bibfield  {author} {\bibinfo {author} {\bibfnamefont {B.~P.}\ \bibnamefont
  {Abbott}} \emph {et~al.} (\bibinfo {collaboration} {Virgo, LIGO
  Scientific}),\ }\href {\doibase 10.1103/PhysRevLett.116.061102} {\bibfield
  {journal} {\bibinfo  {journal} {Phys. Rev. Lett.}\ }\textbf {\bibinfo
  {volume} {116}},\ \bibinfo {pages} {061102} (\bibinfo {year}
  {2016}{\natexlab{a}})},\ \Eprint {http://arxiv.org/abs/1602.03837}
  {arXiv:1602.03837 [gr-qc]} \BibitemShut {NoStop}%
\bibitem [{\citenamefont {Abbott}\ \emph
  {et~al.}(2016{\natexlab{b}})\citenamefont {Abbott} \emph
  {et~al.}}]{Abbott:2016nmj}%
  \BibitemOpen
  \bibfield  {author} {\bibinfo {author} {\bibfnamefont {B.~P.}\ \bibnamefont
  {Abbott}} \emph {et~al.} (\bibinfo {collaboration} {Virgo, LIGO
  Scientific}),\ }\href {\doibase 10.1103/PhysRevLett.116.241103} {\bibfield
  {journal} {\bibinfo  {journal} {Phys. Rev. Lett.}\ }\textbf {\bibinfo
  {volume} {116}},\ \bibinfo {pages} {241103} (\bibinfo {year}
  {2016}{\natexlab{b}})},\ \Eprint {http://arxiv.org/abs/1606.04855}
  {arXiv:1606.04855 [gr-qc]} \BibitemShut {NoStop}%
\bibitem [{\citenamefont {Abbott}\ \emph
  {et~al.}(2016{\natexlab{c}})\citenamefont {Abbott} \emph
  {et~al.}}]{TheLIGOScientific:2016pea}%
  \BibitemOpen
  \bibfield  {author} {\bibinfo {author} {\bibfnamefont {B.~P.}\ \bibnamefont
  {Abbott}} \emph {et~al.} (\bibinfo {collaboration} {Virgo, LIGO
  Scientific}),\ }\href@noop {} {\  (\bibinfo {year} {2016}{\natexlab{c}})},\
  \Eprint {http://arxiv.org/abs/1606.04856} {arXiv:1606.04856 [gr-qc]}
  \BibitemShut {NoStop}%
\bibitem [{\citenamefont {Abbott}\ \emph
  {et~al.}(2016{\natexlab{d}})\citenamefont {Abbott} \emph
  {et~al.}}]{TheLIGOScientific:2016qqj}%
  \BibitemOpen
  \bibfield  {author} {\bibinfo {author} {\bibfnamefont {B.~P.}\ \bibnamefont
  {Abbott}} \emph {et~al.} (\bibinfo {collaboration} {Virgo, LIGO
  Scientific}),\ }\href {\doibase 10.1103/PhysRevD.93.122003} {\bibfield
  {journal} {\bibinfo  {journal} {Phys. Rev.}\ }\textbf {\bibinfo {volume}
  {D93}},\ \bibinfo {pages} {122003} (\bibinfo {year} {2016}{\natexlab{d}})},\
  \Eprint {http://arxiv.org/abs/1602.03839} {arXiv:1602.03839 [gr-qc]}
  \BibitemShut {NoStop}%
\bibitem [{\citenamefont {Buonanno}\ \emph {et~al.}(2007)\citenamefont
  {Buonanno}, \citenamefont {Cook},\ and\ \citenamefont
  {Pretorius}}]{Buonanno:2006ui}%
  \BibitemOpen
  \bibfield  {author} {\bibinfo {author} {\bibfnamefont {A.}~\bibnamefont
  {Buonanno}}, \bibinfo {author} {\bibfnamefont {G.~B.}\ \bibnamefont {Cook}},
  \ and\ \bibinfo {author} {\bibfnamefont {F.}~\bibnamefont {Pretorius}},\
  }\href {\doibase 10.1103/PhysRevD.75.124018} {\bibfield  {journal} {\bibinfo
  {journal} {Phys. Rev.}\ }\textbf {\bibinfo {volume} {D75}},\ \bibinfo {pages}
  {124018} (\bibinfo {year} {2007})},\ \Eprint
  {http://arxiv.org/abs/gr-qc/0610122} {arXiv:gr-qc/0610122 [gr-qc]}
  \BibitemShut {NoStop}%
\bibitem [{\citenamefont {Cardoso}\ \emph {et~al.}(2016)\citenamefont
  {Cardoso}, \citenamefont {Franzin},\ and\ \citenamefont
  {Pani}}]{Cardoso:2016rao}%
  \BibitemOpen
  \bibfield  {author} {\bibinfo {author} {\bibfnamefont {V.}~\bibnamefont
  {Cardoso}}, \bibinfo {author} {\bibfnamefont {E.}~\bibnamefont {Franzin}}, \
  and\ \bibinfo {author} {\bibfnamefont {P.}~\bibnamefont {Pani}},\ }\href
  {\doibase 10.1103/PhysRevLett.116.171101} {\bibfield  {journal} {\bibinfo
  {journal} {Phys. Rev. Lett.}\ }\textbf {\bibinfo {volume} {116}},\ \bibinfo
  {pages} {171101} (\bibinfo {year} {2016})},\ \Eprint
  {http://arxiv.org/abs/1602.07309} {arXiv:1602.07309 [gr-qc]} \BibitemShut
  {NoStop}%
\bibitem [{\citenamefont {Teo}(2003)}]{Teo:2003}%
  \BibitemOpen
  \bibfield  {author} {\bibinfo {author} {\bibfnamefont {E.}~\bibnamefont
  {Teo}},\ }\href@noop {} {\bibfield  {journal} {\bibinfo  {journal}
  {Gen.~Rel.~Gravit.}\ }\textbf {\bibinfo {volume} {35}},\ \bibinfo {pages}
  {1909} (\bibinfo {year} {2003})}\BibitemShut {NoStop}%
\bibitem [{\citenamefont {Hod}(2013)}]{Hod:2012ax}%
  \BibitemOpen
  \bibfield  {author} {\bibinfo {author} {\bibfnamefont {S.}~\bibnamefont
  {Hod}},\ }\href {\doibase 10.1016/j.physletb.2012.12.047} {\bibfield
  {journal} {\bibinfo  {journal} {Phys. Lett.}\ }\textbf {\bibinfo {volume}
  {B718}},\ \bibinfo {pages} {1552} (\bibinfo {year} {2013})},\ \Eprint
  {http://arxiv.org/abs/1210.2486} {arXiv:1210.2486 [gr-qc]} \BibitemShut
  {NoStop}%
\bibitem [{\citenamefont {Goebel}(1972)}]{Goebel:1972}%
  \BibitemOpen
  \bibfield  {author} {\bibinfo {author} {\bibfnamefont {C.~J.}\ \bibnamefont
  {Goebel}},\ }\href@noop {} {\bibfield  {journal} {\bibinfo  {journal}
  {Astrophys.~J.}\ }\textbf {\bibinfo {volume} {172}},\ \bibinfo {pages} {L105}
  (\bibinfo {year} {1972})}\BibitemShut {NoStop}%
\bibitem [{\citenamefont {Mashhoon}(1985)}]{Mashhoon:1985cya}%
  \BibitemOpen
  \bibfield  {author} {\bibinfo {author} {\bibfnamefont {B.}~\bibnamefont
  {Mashhoon}},\ }\href {\doibase 10.1103/PhysRevD.31.290} {\bibfield  {journal}
  {\bibinfo  {journal} {Phys. Rev.}\ }\textbf {\bibinfo {volume} {D31}},\
  \bibinfo {pages} {290} (\bibinfo {year} {1985})}\BibitemShut {NoStop}%
\bibitem [{\citenamefont {Cardoso}\ \emph {et~al.}(2009)\citenamefont
  {Cardoso}, \citenamefont {Miranda}, \citenamefont {Berti}, \citenamefont
  {Witek},\ and\ \citenamefont {Zanchin}}]{Cardoso:2008bp}%
  \BibitemOpen
  \bibfield  {author} {\bibinfo {author} {\bibfnamefont {V.}~\bibnamefont
  {Cardoso}}, \bibinfo {author} {\bibfnamefont {A.~S.}\ \bibnamefont
  {Miranda}}, \bibinfo {author} {\bibfnamefont {E.}~\bibnamefont {Berti}},
  \bibinfo {author} {\bibfnamefont {H.}~\bibnamefont {Witek}}, \ and\ \bibinfo
  {author} {\bibfnamefont {V.~T.}\ \bibnamefont {Zanchin}},\ }\href {\doibase
  10.1103/PhysRevD.79.064016} {\bibfield  {journal} {\bibinfo  {journal} {Phys.
  Rev.}\ }\textbf {\bibinfo {volume} {D79}},\ \bibinfo {pages} {064016}
  (\bibinfo {year} {2009})},\ \Eprint {http://arxiv.org/abs/0812.1806}
  {arXiv:0812.1806 [hep-th]} \BibitemShut {NoStop}%
\bibitem [{\citenamefont {Dolan}(2010)}]{Dolan:2010wr}%
  \BibitemOpen
  \bibfield  {author} {\bibinfo {author} {\bibfnamefont {S.~R.}\ \bibnamefont
  {Dolan}},\ }\href {\doibase 10.1103/PhysRevD.82.104003} {\bibfield  {journal}
  {\bibinfo  {journal} {Phys. Rev.}\ }\textbf {\bibinfo {volume} {D82}},\
  \bibinfo {pages} {104003} (\bibinfo {year} {2010})},\ \Eprint
  {http://arxiv.org/abs/1007.5097} {arXiv:1007.5097 [gr-qc]} \BibitemShut
  {NoStop}%
\bibitem [{\citenamefont {Yang}\ \emph {et~al.}(2012)\citenamefont {Yang},
  \citenamefont {Nichols}, \citenamefont {Zhang}, \citenamefont {Zimmerman},
  \citenamefont {Zhang},\ and\ \citenamefont {Chen}}]{Yang:2012he}%
  \BibitemOpen
  \bibfield  {author} {\bibinfo {author} {\bibfnamefont {H.}~\bibnamefont
  {Yang}}, \bibinfo {author} {\bibfnamefont {D.~A.}\ \bibnamefont {Nichols}},
  \bibinfo {author} {\bibfnamefont {F.}~\bibnamefont {Zhang}}, \bibinfo
  {author} {\bibfnamefont {A.}~\bibnamefont {Zimmerman}}, \bibinfo {author}
  {\bibfnamefont {Z.}~\bibnamefont {Zhang}}, \ and\ \bibinfo {author}
  {\bibfnamefont {Y.}~\bibnamefont {Chen}},\ }\href {\doibase
  10.1103/PhysRevD.86.104006} {\bibfield  {journal} {\bibinfo  {journal} {Phys.
  Rev.}\ }\textbf {\bibinfo {volume} {D86}},\ \bibinfo {pages} {104006}
  (\bibinfo {year} {2012})},\ \Eprint {http://arxiv.org/abs/1207.4253}
  {arXiv:1207.4253 [gr-qc]} \BibitemShut {NoStop}%
\bibitem [{\citenamefont {Perlick}(2004)}]{Perlick:2004tq}%
  \BibitemOpen
  \bibfield  {author} {\bibinfo {author} {\bibfnamefont {V.}~\bibnamefont
  {Perlick}},\ }\href@noop {} {\bibfield  {journal} {\bibinfo  {journal}
  {Living Rev. Rel.}\ }\textbf {\bibinfo {volume} {7}},\ \bibinfo {pages} {9}
  (\bibinfo {year} {2004})}\BibitemShut {NoStop}%
\bibitem [{\citenamefont {Crispino}\ \emph {et~al.}(2009)\citenamefont
  {Crispino}, \citenamefont {Dolan},\ and\ \citenamefont
  {Oliveira}}]{Crispino:2009xt}%
  \BibitemOpen
  \bibfield  {author} {\bibinfo {author} {\bibfnamefont {L.~C.~B.}\
  \bibnamefont {Crispino}}, \bibinfo {author} {\bibfnamefont {S.~R.}\
  \bibnamefont {Dolan}}, \ and\ \bibinfo {author} {\bibfnamefont {E.~S.}\
  \bibnamefont {Oliveira}},\ }\href {\doibase 10.1103/PhysRevLett.102.231103}
  {\bibfield  {journal} {\bibinfo  {journal} {Phys. Rev. Lett.}\ }\textbf
  {\bibinfo {volume} {102}},\ \bibinfo {pages} {231103} (\bibinfo {year}
  {2009})},\ \Eprint {http://arxiv.org/abs/0905.3339} {arXiv:0905.3339 [gr-qc]}
  \BibitemShut {NoStop}%
\bibitem [{\citenamefont {Konoplya}\ and\ \citenamefont
  {Zhidenko}(2011)}]{Konoplya:2011qq}%
  \BibitemOpen
  \bibfield  {author} {\bibinfo {author} {\bibfnamefont {R.~A.}\ \bibnamefont
  {Konoplya}}\ and\ \bibinfo {author} {\bibfnamefont {A.}~\bibnamefont
  {Zhidenko}},\ }\href {\doibase 10.1103/RevModPhys.83.793} {\bibfield
  {journal} {\bibinfo  {journal} {Rev. Mod. Phys.}\ }\textbf {\bibinfo {volume}
  {83}},\ \bibinfo {pages} {793} (\bibinfo {year} {2011})},\ \Eprint
  {http://arxiv.org/abs/1102.4014} {arXiv:1102.4014 [gr-qc]} \BibitemShut
  {NoStop}%
\bibitem [{\citenamefont {Liebling}\ and\ \citenamefont
  {Palenzuela}(2012)}]{Liebling:2012fv}%
  \BibitemOpen
  \bibfield  {author} {\bibinfo {author} {\bibfnamefont {S.~L.}\ \bibnamefont
  {Liebling}}\ and\ \bibinfo {author} {\bibfnamefont {C.}~\bibnamefont
  {Palenzuela}},\ }\href {\doibase 10.12942/lrr-2012-6} {\bibfield  {journal}
  {\bibinfo  {journal} {Living Rev. Rel.}\ }\textbf {\bibinfo {volume} {15}},\
  \bibinfo {pages} {6} (\bibinfo {year} {2012})},\ \Eprint
  {http://arxiv.org/abs/1202.5809} {arXiv:1202.5809 [gr-qc]} \BibitemShut
  {NoStop}%
\bibitem [{\citenamefont {Chirenti}\ and\ \citenamefont
  {Rezzolla}(2016)}]{Chirenti:2016hzd}%
  \BibitemOpen
  \bibfield  {author} {\bibinfo {author} {\bibfnamefont {C.}~\bibnamefont
  {Chirenti}}\ and\ \bibinfo {author} {\bibfnamefont {L.}~\bibnamefont
  {Rezzolla}},\ }\href@noop {} {\  (\bibinfo {year} {2016})},\ \Eprint
  {http://arxiv.org/abs/1602.08759} {arXiv:1602.08759 [gr-qc]} \BibitemShut
  {NoStop}%
\bibitem [{\citenamefont {Damour}\ and\ \citenamefont
  {Solodukhin}(2007)}]{Damour:2007ap}%
  \BibitemOpen
  \bibfield  {author} {\bibinfo {author} {\bibfnamefont {T.}~\bibnamefont
  {Damour}}\ and\ \bibinfo {author} {\bibfnamefont {S.~N.}\ \bibnamefont
  {Solodukhin}},\ }\href {\doibase 10.1103/PhysRevD.76.024016} {\bibfield
  {journal} {\bibinfo  {journal} {Phys. Rev.}\ }\textbf {\bibinfo {volume}
  {D76}},\ \bibinfo {pages} {024016} (\bibinfo {year} {2007})},\ \Eprint
  {http://arxiv.org/abs/0704.2667} {arXiv:0704.2667 [gr-qc]} \BibitemShut
  {NoStop}%
\bibitem [{\citenamefont {Cardoso}\ \emph {et~al.}(2014)\citenamefont
  {Cardoso}, \citenamefont {Crispino}, \citenamefont {Macedo}, \citenamefont
  {Okawa},\ and\ \citenamefont {Pani}}]{Cardoso:2014sna}%
  \BibitemOpen
  \bibfield  {author} {\bibinfo {author} {\bibfnamefont {V.}~\bibnamefont
  {Cardoso}}, \bibinfo {author} {\bibfnamefont {L.~C.~B.}\ \bibnamefont
  {Crispino}}, \bibinfo {author} {\bibfnamefont {C.~F.~B.}\ \bibnamefont
  {Macedo}}, \bibinfo {author} {\bibfnamefont {H.}~\bibnamefont {Okawa}}, \
  and\ \bibinfo {author} {\bibfnamefont {P.}~\bibnamefont {Pani}},\ }\href
  {\doibase 10.1103/PhysRevD.90.044069} {\bibfield  {journal} {\bibinfo
  {journal} {Phys. Rev.}\ }\textbf {\bibinfo {volume} {D90}},\ \bibinfo {pages}
  {044069} (\bibinfo {year} {2014})},\ \Eprint {http://arxiv.org/abs/1406.5510}
  {arXiv:1406.5510 [gr-qc]} \BibitemShut {NoStop}%
\bibitem [{\citenamefont {Friedman}(1978)}]{Friedman:1978}%
  \BibitemOpen
  \bibfield  {author} {\bibinfo {author} {\bibfnamefont {J.~L.}\ \bibnamefont
  {Friedman}},\ }\href@noop {} {\bibfield  {journal} {\bibinfo  {journal}
  {Commun.~Math.~Phys.}\ }\textbf {\bibinfo {volume} {63}},\ \bibinfo {pages}
  {243} (\bibinfo {year} {1978})}\BibitemShut {NoStop}%
\bibitem [{\citenamefont {Pani}\ \emph {et~al.}(2010)\citenamefont {Pani},
  \citenamefont {Barausse}, \citenamefont {Berti},\ and\ \citenamefont
  {Cardoso}}]{Pani:2010jz}%
  \BibitemOpen
  \bibfield  {author} {\bibinfo {author} {\bibfnamefont {P.}~\bibnamefont
  {Pani}}, \bibinfo {author} {\bibfnamefont {E.}~\bibnamefont {Barausse}},
  \bibinfo {author} {\bibfnamefont {E.}~\bibnamefont {Berti}}, \ and\ \bibinfo
  {author} {\bibfnamefont {V.}~\bibnamefont {Cardoso}},\ }\href {\doibase
  10.1103/PhysRevD.82.044009} {\bibfield  {journal} {\bibinfo  {journal} {Phys.
  Rev.}\ }\textbf {\bibinfo {volume} {D82}},\ \bibinfo {pages} {044009}
  (\bibinfo {year} {2010})},\ \Eprint {http://arxiv.org/abs/1006.1863}
  {arXiv:1006.1863 [gr-qc]} \BibitemShut {NoStop}%
\bibitem [{\citenamefont {Keir}(2016)}]{Keir:2014oka}%
  \BibitemOpen
  \bibfield  {author} {\bibinfo {author} {\bibfnamefont {J.}~\bibnamefont
  {Keir}},\ }\href {\doibase 10.1088/0264-9381/33/13/135009} {\bibfield
  {journal} {\bibinfo  {journal} {Class. Quant. Grav.}\ }\textbf {\bibinfo
  {volume} {33}},\ \bibinfo {pages} {135009} (\bibinfo {year} {2016})},\
  \Eprint {http://arxiv.org/abs/1404.7036} {arXiv:1404.7036 [gr-qc]}
  \BibitemShut {NoStop}%
\bibitem [{\citenamefont {Casals}\ and\ \citenamefont
  {Ottewill}(2015)}]{Casals:2015nja}%
  \BibitemOpen
  \bibfield  {author} {\bibinfo {author} {\bibfnamefont {M.}~\bibnamefont
  {Casals}}\ and\ \bibinfo {author} {\bibfnamefont {A.~C.}\ \bibnamefont
  {Ottewill}},\ }\href {\doibase 10.1103/PhysRevD.92.124055} {\bibfield
  {journal} {\bibinfo  {journal} {Phys. Rev.}\ }\textbf {\bibinfo {volume}
  {D92}},\ \bibinfo {pages} {124055} (\bibinfo {year} {2015})},\ \Eprint
  {http://arxiv.org/abs/1509.04702} {arXiv:1509.04702 [gr-qc]} \BibitemShut
  {NoStop}%
\bibitem [{\citenamefont {Shipley}\ and\ \citenamefont
  {Dolan}(2016)}]{Shipley:2016omi}%
  \BibitemOpen
  \bibfield  {author} {\bibinfo {author} {\bibfnamefont {J.~O.}\ \bibnamefont
  {Shipley}}\ and\ \bibinfo {author} {\bibfnamefont {S.~R.}\ \bibnamefont
  {Dolan}},\ }\href@noop {} {\  (\bibinfo {year} {2016})},\ \Eprint
  {http://arxiv.org/abs/1603.04469} {arXiv:1603.04469 [gr-qc]} \BibitemShut
  {NoStop}%
\bibitem [{\citenamefont {Cunha}\ \emph {et~al.}(2015)\citenamefont {Cunha},
  \citenamefont {Herdeiro}, \citenamefont {Radu},\ and\ \citenamefont
  {Runarsson}}]{Cunha:2015yba}%
  \BibitemOpen
  \bibfield  {author} {\bibinfo {author} {\bibfnamefont {P.~V.~P.}\
  \bibnamefont {Cunha}}, \bibinfo {author} {\bibfnamefont {C.~A.~R.}\
  \bibnamefont {Herdeiro}}, \bibinfo {author} {\bibfnamefont {E.}~\bibnamefont
  {Radu}}, \ and\ \bibinfo {author} {\bibfnamefont {H.~F.}\ \bibnamefont
  {Runarsson}},\ }\href {\doibase 10.1103/PhysRevLett.115.211102} {\bibfield
  {journal} {\bibinfo  {journal} {Phys. Rev. Lett.}\ }\textbf {\bibinfo
  {volume} {115}},\ \bibinfo {pages} {211102} (\bibinfo {year} {2015})},\
  \Eprint {http://arxiv.org/abs/1509.00021} {arXiv:1509.00021 [gr-qc]}
  \BibitemShut {NoStop}%
\bibitem [{\citenamefont {Liang}(1974)}]{Liang:1974}%
  \BibitemOpen
  \bibfield  {author} {\bibinfo {author} {\bibfnamefont {E.~P.~T.}\
  \bibnamefont {Liang}},\ }\href {\doibase 10.1103/PhysRevD.9.3257} {\bibfield
  {journal} {\bibinfo  {journal} {Phys. Rev. D}\ }\textbf {\bibinfo {volume}
  {9}},\ \bibinfo {pages} {3257} (\bibinfo {year} {1974})}\BibitemShut
  {NoStop}%
\bibitem [{\citenamefont {Calvani}\ \emph {et~al.}(1980)\citenamefont
  {Calvani}, \citenamefont {De~Felice},\ and\ \citenamefont
  {Nobili}}]{Calvani:1980is}%
  \BibitemOpen
  \bibfield  {author} {\bibinfo {author} {\bibfnamefont {M.}~\bibnamefont
  {Calvani}}, \bibinfo {author} {\bibfnamefont {F.}~\bibnamefont {De~Felice}},
  \ and\ \bibinfo {author} {\bibfnamefont {L.}~\bibnamefont {Nobili}},\ }\href
  {\doibase 10.1088/0305-4470/13/10/018} {\bibfield  {journal} {\bibinfo
  {journal} {J. Phys.}\ }\textbf {\bibinfo {volume} {A13}},\ \bibinfo {pages}
  {3213} (\bibinfo {year} {1980})}\BibitemShut {NoStop}%
\bibitem [{\citenamefont {Stuchl{\'\i}k}(1981)}]{Stuchlik:1981}%
  \BibitemOpen
  \bibfield  {author} {\bibinfo {author} {\bibfnamefont {Z.}~\bibnamefont
  {Stuchl{\'\i}k}},\ }\href@noop {} {\bibfield  {journal} {\bibinfo  {journal}
  {Bulletin of the Astronomical Institutes of Czechoslovakia}\ }\textbf
  {\bibinfo {volume} {32}},\ \bibinfo {pages} {366} (\bibinfo {year}
  {1981})}\BibitemShut {NoStop}%
\bibitem [{\citenamefont {Balek}\ \emph {et~al.}(1989)\citenamefont {Balek},
  \citenamefont {Bicak},\ and\ \citenamefont {Stuchlik}}]{Balek:1989}%
  \BibitemOpen
  \bibfield  {author} {\bibinfo {author} {\bibfnamefont {V.}~\bibnamefont
  {Balek}}, \bibinfo {author} {\bibfnamefont {J.}~\bibnamefont {Bicak}}, \ and\
  \bibinfo {author} {\bibfnamefont {Z.}~\bibnamefont {Stuchlik}},\ }\href@noop
  {} {\bibfield  {journal} {\bibinfo  {journal} {Bulletin of the Astronomical
  Institutes of Czechoslovakia}\ }\textbf {\bibinfo {volume} {40}},\ \bibinfo
  {pages} {133} (\bibinfo {year} {1989})}\BibitemShut {NoStop}%
\bibitem [{\citenamefont {Pugliese}\ \emph {et~al.}(2013)\citenamefont
  {Pugliese}, \citenamefont {Quevedo},\ and\ \citenamefont
  {Ruffini}}]{Pugliese:2013zma}%
  \BibitemOpen
  \bibfield  {author} {\bibinfo {author} {\bibfnamefont {D.}~\bibnamefont
  {Pugliese}}, \bibinfo {author} {\bibfnamefont {H.}~\bibnamefont {Quevedo}}, \
  and\ \bibinfo {author} {\bibfnamefont {R.}~\bibnamefont {Ruffini}},\ }\href
  {\doibase 10.1103/PhysRevD.88.024042} {\bibfield  {journal} {\bibinfo
  {journal} {Phys. Rev.}\ }\textbf {\bibinfo {volume} {D88}},\ \bibinfo {pages}
  {024042} (\bibinfo {year} {2013})},\ \Eprint {http://arxiv.org/abs/1303.6250}
  {arXiv:1303.6250 [gr-qc]} \BibitemShut {NoStop}%
\bibitem [{\citenamefont {Dokuchaev}(2011)}]{Dokuchaev:2011wm}%
  \BibitemOpen
  \bibfield  {author} {\bibinfo {author} {\bibfnamefont {V.~I.}\ \bibnamefont
  {Dokuchaev}},\ }\href {\doibase 10.1088/0264-9381/28/23/235015} {\bibfield
  {journal} {\bibinfo  {journal} {Class. Quant. Grav.}\ }\textbf {\bibinfo
  {volume} {28}},\ \bibinfo {pages} {235015} (\bibinfo {year} {2011})},\
  \Eprint {http://arxiv.org/abs/1103.6140} {arXiv:1103.6140 [gr-qc]}
  \BibitemShut {NoStop}%
\bibitem [{\citenamefont {Ulbricht}\ and\ \citenamefont
  {Meinel}(2015)}]{Ulbricht:2015vwa}%
  \BibitemOpen
  \bibfield  {author} {\bibinfo {author} {\bibfnamefont {S.}~\bibnamefont
  {Ulbricht}}\ and\ \bibinfo {author} {\bibfnamefont {R.}~\bibnamefont
  {Meinel}},\ }\href {\doibase 10.1088/0264-9381/32/14/147001} {\bibfield
  {journal} {\bibinfo  {journal} {Class. Quant. Grav.}\ }\textbf {\bibinfo
  {volume} {32}},\ \bibinfo {pages} {147001} (\bibinfo {year} {2015})},\
  \Eprint {http://arxiv.org/abs/1503.01973} {arXiv:1503.01973 [gr-qc]}
  \BibitemShut {NoStop}%
\bibitem [{\citenamefont {Khoo}\ and\ \citenamefont
  {Ong}(2016)}]{Khoo:2016xqv}%
  \BibitemOpen
  \bibfield  {author} {\bibinfo {author} {\bibfnamefont {F.~S.}\ \bibnamefont
  {Khoo}}\ and\ \bibinfo {author} {\bibfnamefont {Y.~C.}\ \bibnamefont {Ong}},\
  }\href@noop {} {\  (\bibinfo {year} {2016})},\ \Eprint
  {http://arxiv.org/abs/1605.05774} {arXiv:1605.05774 [gr-qc]} \BibitemShut
  {NoStop}%
\bibitem [{\citenamefont {Stuchl\'ik}\ and\ \citenamefont
  {Hled\'ik}(2002)}]{Stuchlik:2002}%
  \BibitemOpen
  \bibfield  {author} {\bibinfo {author} {\bibfnamefont {Z.}~\bibnamefont
  {Stuchl\'ik}}\ and\ \bibinfo {author} {\bibfnamefont {S.}~\bibnamefont
  {Hled\'ik}},\ }\href@noop {} {\bibfield  {journal} {\bibinfo  {journal} {Acta
  Physica Slovaca}\ }\textbf {\bibinfo {volume} {52}},\ \bibinfo {pages} {363}
  (\bibinfo {year} {2002})},\ \Eprint {http://arxiv.org/abs/0803.2685}
  {arXiv:0803.2685 [gr-qc]} \BibitemShut {NoStop}%
\bibitem [{\citenamefont {Grenzebach}\ \emph {et~al.}(2014)\citenamefont
  {Grenzebach}, \citenamefont {Perlick},\ and\ \citenamefont
  {Lammerzahl}}]{Grenzebach:2014fha}%
  \BibitemOpen
  \bibfield  {author} {\bibinfo {author} {\bibfnamefont {A.}~\bibnamefont
  {Grenzebach}}, \bibinfo {author} {\bibfnamefont {V.}~\bibnamefont {Perlick}},
  \ and\ \bibinfo {author} {\bibfnamefont {C.}~\bibnamefont {Lammerzahl}},\
  }\href {\doibase 10.1103/PhysRevD.89.124004} {\bibfield  {journal} {\bibinfo
  {journal} {Phys. Rev.}\ }\textbf {\bibinfo {volume} {D89}},\ \bibinfo {pages}
  {124004} (\bibinfo {year} {2014})},\ \Eprint {http://arxiv.org/abs/1403.5234}
  {arXiv:1403.5234 [gr-qc]} \BibitemShut {NoStop}%
\bibitem [{\citenamefont {Igata}\ \emph {et~al.}(2013)\citenamefont {Igata},
  \citenamefont {Ishihara},\ and\ \citenamefont {Takamori}}]{Igata:2013be}%
  \BibitemOpen
  \bibfield  {author} {\bibinfo {author} {\bibfnamefont {T.}~\bibnamefont
  {Igata}}, \bibinfo {author} {\bibfnamefont {H.}~\bibnamefont {Ishihara}}, \
  and\ \bibinfo {author} {\bibfnamefont {Y.}~\bibnamefont {Takamori}},\ }\href
  {\doibase 10.1103/PhysRevD.87.104005} {\bibfield  {journal} {\bibinfo
  {journal} {Phys. Rev.}\ }\textbf {\bibinfo {volume} {D87}},\ \bibinfo {pages}
  {104005} (\bibinfo {year} {2013})},\ \Eprint {http://arxiv.org/abs/1302.0291}
  {arXiv:1302.0291 [hep-th]} \BibitemShut {NoStop}%
\bibitem [{\citenamefont {Igata}(2015)}]{Igata:2014xca}%
  \BibitemOpen
  \bibfield  {author} {\bibinfo {author} {\bibfnamefont {T.}~\bibnamefont
  {Igata}},\ }\href {\doibase 10.1103/PhysRevD.92.024002} {\bibfield  {journal}
  {\bibinfo  {journal} {Phys. Rev.}\ }\textbf {\bibinfo {volume} {D92}},\
  \bibinfo {pages} {024002} (\bibinfo {year} {2015})},\ \Eprint
  {http://arxiv.org/abs/1411.6102} {arXiv:1411.6102 [gr-qc]} \BibitemShut
  {NoStop}%
\bibitem [{\citenamefont {W\"unsch}\ \emph {et~al.}(2013)\citenamefont
  {W\"unsch}, \citenamefont {M\"uller}, \citenamefont {Weiskopf},\ and\
  \citenamefont {Wunner}}]{Wunsch:2013st}%
  \BibitemOpen
  \bibfield  {author} {\bibinfo {author} {\bibfnamefont {A.}~\bibnamefont
  {W\"unsch}}, \bibinfo {author} {\bibfnamefont {T.}~\bibnamefont {M\"uller}},
  \bibinfo {author} {\bibfnamefont {D.}~\bibnamefont {Weiskopf}}, \ and\
  \bibinfo {author} {\bibfnamefont {G.}~\bibnamefont {Wunner}},\ }\href
  {\doibase 10.1103/PhysRevD.87.024007} {\bibfield  {journal} {\bibinfo
  {journal} {Phys. Rev.}\ }\textbf {\bibinfo {volume} {D87}},\ \bibinfo {pages}
  {024007} (\bibinfo {year} {2013})},\ \Eprint {http://arxiv.org/abs/1301.7560}
  {arXiv:1301.7560 [gr-qc]} \BibitemShut {NoStop}%
\bibitem [{\citenamefont {Robinson}(1975)}]{Robinson:1975bv}%
  \BibitemOpen
  \bibfield  {author} {\bibinfo {author} {\bibfnamefont {D.~C.}\ \bibnamefont
  {Robinson}},\ }\href {\doibase 10.1103/PhysRevLett.34.905} {\bibfield
  {journal} {\bibinfo  {journal} {Phys. Rev. Lett.}\ }\textbf {\bibinfo
  {volume} {34}},\ \bibinfo {pages} {905} (\bibinfo {year} {1975})}\BibitemShut
  {NoStop}%
\bibitem [{\citenamefont {Mazur}(1982)}]{Mazur:1982db}%
  \BibitemOpen
  \bibfield  {author} {\bibinfo {author} {\bibfnamefont {P.~O.}\ \bibnamefont
  {Mazur}},\ }\href {\doibase 10.1088/0305-4470/15/10/021} {\bibfield
  {journal} {\bibinfo  {journal} {J. Phys.}\ }\textbf {\bibinfo {volume}
  {A15}},\ \bibinfo {pages} {3173} (\bibinfo {year} {1982})}\BibitemShut
  {NoStop}%
\bibitem [{\citenamefont {Carter}(1968)}]{Carter:1968rr}%
  \BibitemOpen
  \bibfield  {author} {\bibinfo {author} {\bibfnamefont {B.}~\bibnamefont
  {Carter}},\ }\href {\doibase 10.1103/PhysRev.174.1559} {\bibfield  {journal}
  {\bibinfo  {journal} {Phys. Rev.}\ }\textbf {\bibinfo {volume} {174}},\
  \bibinfo {pages} {1559} (\bibinfo {year} {1968})}\BibitemShut {NoStop}%
\bibitem [{\citenamefont {Stephani}\ \emph {et~al.}(2004)\citenamefont
  {Stephani}, \citenamefont {Kramer}, \citenamefont {MacCallum}, \citenamefont
  {Hoenselaers},\ and\ \citenamefont {Herlt}}]{Stephani:2003tm}%
  \BibitemOpen
  \bibfield  {author} {\bibinfo {author} {\bibfnamefont {H.}~\bibnamefont
  {Stephani}}, \bibinfo {author} {\bibfnamefont {D.}~\bibnamefont {Kramer}},
  \bibinfo {author} {\bibfnamefont {M.~A.~H.}\ \bibnamefont {MacCallum}},
  \bibinfo {author} {\bibfnamefont {C.}~\bibnamefont {Hoenselaers}}, \ and\
  \bibinfo {author} {\bibfnamefont {E.}~\bibnamefont {Herlt}},\ }\href
  {http://www.cambridge.org/uk/catalogue/catalogue.asp?isbn=0521461367} {\emph
  {\bibinfo {title} {{Exact solutions of Einstein's field equations}}}}\
  (\bibinfo  {publisher} {Cambridge University Press},\ \bibinfo {year}
  {2004})\BibitemShut {NoStop}%
\bibitem [{\citenamefont {Majumdar}(1947)}]{Majumdar:1947eu}%
  \BibitemOpen
  \bibfield  {author} {\bibinfo {author} {\bibfnamefont {S.~D.}\ \bibnamefont
  {Majumdar}},\ }\href {\doibase 10.1103/PhysRev.72.390} {\bibfield  {journal}
  {\bibinfo  {journal} {Phys. Rev.}\ }\textbf {\bibinfo {volume} {72}},\
  \bibinfo {pages} {390} (\bibinfo {year} {1947})}\BibitemShut {NoStop}%
\bibitem [{\citenamefont {Papapetrou}(1945)}]{Papapetrou:1947}%
  \BibitemOpen
  \bibfield  {author} {\bibinfo {author} {\bibfnamefont {A.}~\bibnamefont
  {Papapetrou}},\ }\href {http://www.jstor.org/stable/20488481} {\bibfield
  {journal} {\bibinfo  {journal} {P.~Roy.~Irish Acad.~A}\ }\textbf {\bibinfo
  {volume} {51}},\ \bibinfo {pages} {191} (\bibinfo {year} {1945})}\BibitemShut
  {NoStop}%
\bibitem [{\citenamefont {Chandrasekhar}(1989)}]{Chandrasekhar:1989vk}%
  \BibitemOpen
  \bibfield  {author} {\bibinfo {author} {\bibfnamefont {S.}~\bibnamefont
  {Chandrasekhar}},\ }\href {\doibase 10.1098/rspa.1989.0010} {\bibfield
  {journal} {\bibinfo  {journal} {Proc. Roy. Soc. Lond.}\ }\textbf {\bibinfo
  {volume} {A421}},\ \bibinfo {pages} {227} (\bibinfo {year}
  {1989})}\BibitemShut {NoStop}%
\bibitem [{\citenamefont {Hartle}\ and\ \citenamefont
  {Hawking}(1972)}]{Hartle:1972ya}%
  \BibitemOpen
  \bibfield  {author} {\bibinfo {author} {\bibfnamefont {J.~B.}\ \bibnamefont
  {Hartle}}\ and\ \bibinfo {author} {\bibfnamefont {S.~W.}\ \bibnamefont
  {Hawking}},\ }\href {\doibase 10.1007/BF01645696} {\bibfield  {journal}
  {\bibinfo  {journal} {Commun. Math. Phys.}\ }\textbf {\bibinfo {volume}
  {26}},\ \bibinfo {pages} {87} (\bibinfo {year} {1972})}\BibitemShut {NoStop}%
\bibitem [{\citenamefont {Coelho}\ and\ \citenamefont
  {Herdeiro}(2009)}]{CoelhoHerdeiro2009}%
  \BibitemOpen
  \bibfield  {author} {\bibinfo {author} {\bibfnamefont {F.~S.}\ \bibnamefont
  {Coelho}}\ and\ \bibinfo {author} {\bibfnamefont {C.~A.~R.}\ \bibnamefont
  {Herdeiro}},\ }\href {\doibase 10.1103/PhysRevD.80.104036} {\bibfield
  {journal} {\bibinfo  {journal} {Phys. Rev. D}\ }\textbf {\bibinfo {volume}
  {80}},\ \bibinfo {pages} {104036} (\bibinfo {year} {2009})}\BibitemShut
  {NoStop}%
\bibitem [{\citenamefont {Perry}\ and\ \citenamefont
  {Cooperstock}(1997)}]{Perry:1996ja}%
  \BibitemOpen
  \bibfield  {author} {\bibinfo {author} {\bibfnamefont {G.~P.}\ \bibnamefont
  {Perry}}\ and\ \bibinfo {author} {\bibfnamefont {F.~I.}\ \bibnamefont
  {Cooperstock}},\ }\href {\doibase 10.1088/0264-9381/14/5/032} {\bibfield
  {journal} {\bibinfo  {journal} {Class. Quant. Grav.}\ }\textbf {\bibinfo
  {volume} {14}},\ \bibinfo {pages} {1329} (\bibinfo {year} {1997})},\ \Eprint
  {http://arxiv.org/abs/gr-qc/9611066} {arXiv:gr-qc/9611066 [gr-qc]}
  \BibitemShut {NoStop}%
\bibitem [{\citenamefont {Bret\'on}\ \emph {et~al.}(1998)\citenamefont
  {Bret\'on}, \citenamefont {Manko},\ and\ \citenamefont
  {S\'anchez}}]{BMA:1998}%
  \BibitemOpen
  \bibfield  {author} {\bibinfo {author} {\bibfnamefont {N.}~\bibnamefont
  {Bret\'on}}, \bibinfo {author} {\bibfnamefont {V.~S.}\ \bibnamefont {Manko}},
  \ and\ \bibinfo {author} {\bibfnamefont {J.~A.}\ \bibnamefont {S\'anchez}},\
  }\href {http://stacks.iop.org/0264-9381/15/i=10/a=013} {\bibfield  {journal}
  {\bibinfo  {journal} {Classical and Quantum Gravity}\ }\textbf {\bibinfo
  {volume} {15}},\ \bibinfo {pages} {3071} (\bibinfo {year}
  {1998})}\BibitemShut {NoStop}%
\bibitem [{\citenamefont {Varzugin}\ and\ \citenamefont
  {Chistyakov}(2002)}]{Varzugin:2001ab}%
  \BibitemOpen
  \bibfield  {author} {\bibinfo {author} {\bibfnamefont {G.~G.}\ \bibnamefont
  {Varzugin}}\ and\ \bibinfo {author} {\bibfnamefont {A.~S.}\ \bibnamefont
  {Chistyakov}},\ }\href {\doibase 10.1088/0264-9381/19/17/307} {\bibfield
  {journal} {\bibinfo  {journal} {Class. Quant. Grav.}\ }\textbf {\bibinfo
  {volume} {19}},\ \bibinfo {pages} {4553} (\bibinfo {year} {2002})},\ \Eprint
  {http://arxiv.org/abs/gr-qc/0112003} {arXiv:gr-qc/0112003 [gr-qc]}
  \BibitemShut {NoStop}%
\bibitem [{\citenamefont {Alekseev}\ and\ \citenamefont
  {Belinski}(2007)}]{Alekseev:Belinski:2007}%
  \BibitemOpen
  \bibfield  {author} {\bibinfo {author} {\bibfnamefont {G.~A.}\ \bibnamefont
  {Alekseev}}\ and\ \bibinfo {author} {\bibfnamefont {V.~A.}\ \bibnamefont
  {Belinski}},\ }\href {\doibase 10.1103/PhysRevD.76.021501} {\bibfield
  {journal} {\bibinfo  {journal} {Phys. Rev. D}\ }\textbf {\bibinfo {volume}
  {76}},\ \bibinfo {pages} {021501} (\bibinfo {year} {2007})}\BibitemShut
  {NoStop}%
\bibitem [{\citenamefont {Manko}(2007)}]{Manko:2007}%
  \BibitemOpen
  \bibfield  {author} {\bibinfo {author} {\bibfnamefont {V.~S.}\ \bibnamefont
  {Manko}},\ }\href {\doibase 10.1103/PhysRevD.76.124032} {\bibfield  {journal}
  {\bibinfo  {journal} {Phys. Rev. D}\ }\textbf {\bibinfo {volume} {76}},\
  \bibinfo {pages} {124032} (\bibinfo {year} {2007})}\BibitemShut {NoStop}%
\bibitem [{\citenamefont {Bach}\ and\ \citenamefont {Weyl}(2012)}]{Bach:2012}%
  \BibitemOpen
  \bibfield  {author} {\bibinfo {author} {\bibfnamefont {R.}~\bibnamefont
  {Bach}}\ and\ \bibinfo {author} {\bibfnamefont {H.}~\bibnamefont {Weyl}},\
  }\href@noop {} {\bibfield  {journal} {\bibinfo  {journal}
  {Gen.~Rel.~Gravit.}\ }\textbf {\bibinfo {volume} {44}},\ \bibinfo {pages}
  {817} (\bibinfo {year} {2012})}\BibitemShut {NoStop}%
\bibitem [{\citenamefont {Berry}(1978)}]{Berry:1978}%
  \BibitemOpen
  \bibfield  {author} {\bibinfo {author} {\bibfnamefont {M.~V.}\ \bibnamefont
  {Berry}},\ }\href@noop {} {\bibfield  {journal} {\bibinfo  {journal} {Topics
  in Nonlinear Dynamics}\ }\textbf {\bibinfo {volume} {46}},\ \bibinfo {pages}
  {16} (\bibinfo {year} {1978})}\BibitemShut {NoStop}%
\bibitem [{\citenamefont {H{\'e}non}\ and\ \citenamefont
  {Heiles}(1964)}]{Henon:1964}%
  \BibitemOpen
  \bibfield  {author} {\bibinfo {author} {\bibfnamefont {M.}~\bibnamefont
  {H{\'e}non}}\ and\ \bibinfo {author} {\bibfnamefont {C.}~\bibnamefont
  {Heiles}},\ }\href@noop {} {\bibfield  {journal} {\bibinfo  {journal}
  {Astron.~J.}\ }\textbf {\bibinfo {volume} {69}},\ \bibinfo {pages} {73}
  (\bibinfo {year} {1964})}\BibitemShut {NoStop}%
\bibitem [{\citenamefont {H{\'e}non}(1983)}]{Henon:1983}%
  \BibitemOpen
  \bibfield  {author} {\bibinfo {author} {\bibfnamefont {M.}~\bibnamefont
  {H{\'e}non}},\ }\href@noop {} {\bibfield  {journal} {\bibinfo  {journal}
  {Chaotic Behavior of Deterministic Systems (Les Houches, 1981)}\ ,\ \bibinfo
  {pages} {53}} (\bibinfo {year} {1983})}\BibitemShut {NoStop}%
\bibitem [{Note1()}]{Note1}%
  \BibitemOpen
  \bibinfo {note} {This may also be inferred by taking the limit $v\rightarrow
  1$ of Eq.~(32) in Ref.~\cite {Bardeen:1970}}\BibitemShut {NoStop}%
\bibitem [{\citenamefont {Misner}\ \emph {et~al.}(1972)\citenamefont {Misner},
  \citenamefont {Breuer}, \citenamefont {Brill}, \citenamefont {Chrzanowski},
  \citenamefont {Hughes},\ and\ \citenamefont {Pereira}}]{Misner:1972jf}%
  \BibitemOpen
  \bibfield  {author} {\bibinfo {author} {\bibfnamefont {C.~W.}\ \bibnamefont
  {Misner}}, \bibinfo {author} {\bibfnamefont {R.~A.}\ \bibnamefont {Breuer}},
  \bibinfo {author} {\bibfnamefont {D.~R.}\ \bibnamefont {Brill}}, \bibinfo
  {author} {\bibfnamefont {P.~L.}\ \bibnamefont {Chrzanowski}}, \bibinfo
  {author} {\bibfnamefont {H.~G.}\ \bibnamefont {Hughes}}, \ and\ \bibinfo
  {author} {\bibfnamefont {C.~M.}\ \bibnamefont {Pereira}},\ }\href {\doibase
  10.1103/PhysRevLett.28.998} {\bibfield  {journal} {\bibinfo  {journal} {Phys.
  Rev. Lett.}\ }\textbf {\bibinfo {volume} {28}},\ \bibinfo {pages} {998}
  (\bibinfo {year} {1972})}\BibitemShut {NoStop}%
\bibitem [{\citenamefont {Breuer}\ \emph {et~al.}(1973)\citenamefont {Breuer},
  \citenamefont {Chrzanowksi}, \citenamefont {Hughes},\ and\ \citenamefont
  {Misner}}]{Breuer:1973}%
  \BibitemOpen
  \bibfield  {author} {\bibinfo {author} {\bibfnamefont {R.~A.}\ \bibnamefont
  {Breuer}}, \bibinfo {author} {\bibfnamefont {P.~L.}\ \bibnamefont
  {Chrzanowksi}}, \bibinfo {author} {\bibfnamefont {H.~G.}\ \bibnamefont
  {Hughes}}, \ and\ \bibinfo {author} {\bibfnamefont {C.~W.}\ \bibnamefont
  {Misner}},\ }\href {\doibase 10.1103/PhysRevD.8.4309} {\bibfield  {journal}
  {\bibinfo  {journal} {Phys. Rev. D}\ }\textbf {\bibinfo {volume} {8}},\
  \bibinfo {pages} {4309} (\bibinfo {year} {1973})}\BibitemShut {NoStop}%
\bibitem [{\citenamefont {Chrzanowski}\ and\ \citenamefont
  {Misner}(1974)}]{Chrzanowski:1974nr}%
  \BibitemOpen
  \bibfield  {author} {\bibinfo {author} {\bibfnamefont {P.~L.}\ \bibnamefont
  {Chrzanowski}}\ and\ \bibinfo {author} {\bibfnamefont {C.~W.}\ \bibnamefont
  {Misner}},\ }\href {\doibase 10.1103/PhysRevD.10.1701} {\bibfield  {journal}
  {\bibinfo  {journal} {Phys. Rev.}\ }\textbf {\bibinfo {volume} {D10}},\
  \bibinfo {pages} {1701} (\bibinfo {year} {1974})}\BibitemShut {NoStop}%
\bibitem [{\citenamefont {Misner}(1972)}]{Misner:1972kx}%
  \BibitemOpen
  \bibfield  {author} {\bibinfo {author} {\bibfnamefont {C.~W.}\ \bibnamefont
  {Misner}},\ }\href {\doibase 10.1103/PhysRevLett.28.994} {\bibfield
  {journal} {\bibinfo  {journal} {Phys. Rev. Lett.}\ }\textbf {\bibinfo
  {volume} {28}},\ \bibinfo {pages} {994} (\bibinfo {year} {1972})}\BibitemShut
  {NoStop}%
\bibitem [{\citenamefont {Davis}\ \emph {et~al.}(1972)\citenamefont {Davis},
  \citenamefont {Ruffini}, \citenamefont {Tiomno},\ and\ \citenamefont
  {Zerilli}}]{Davis:1972dm}%
  \BibitemOpen
  \bibfield  {author} {\bibinfo {author} {\bibfnamefont {M.}~\bibnamefont
  {Davis}}, \bibinfo {author} {\bibfnamefont {R.}~\bibnamefont {Ruffini}},
  \bibinfo {author} {\bibfnamefont {J.}~\bibnamefont {Tiomno}}, \ and\ \bibinfo
  {author} {\bibfnamefont {F.}~\bibnamefont {Zerilli}},\ }\href {\doibase
  10.1103/PhysRevLett.28.1352} {\bibfield  {journal} {\bibinfo  {journal}
  {Phys. Rev. Lett.}\ }\textbf {\bibinfo {volume} {28}},\ \bibinfo {pages}
  {1352} (\bibinfo {year} {1972})}\BibitemShut {NoStop}%
\bibitem [{\citenamefont {Chitre}\ and\ \citenamefont
  {Price}(1972)}]{Chitre:1972fv}%
  \BibitemOpen
  \bibfield  {author} {\bibinfo {author} {\bibfnamefont {D.~M.}\ \bibnamefont
  {Chitre}}\ and\ \bibinfo {author} {\bibfnamefont {R.~H.}\ \bibnamefont
  {Price}},\ }\href {\doibase 10.1103/PhysRevLett.29.185} {\bibfield  {journal}
  {\bibinfo  {journal} {Phys. Rev. Lett.}\ }\textbf {\bibinfo {volume} {29}},\
  \bibinfo {pages} {185} (\bibinfo {year} {1972})}\BibitemShut {NoStop}%
\bibitem [{\citenamefont {Ernst}(1968)}]{Ernst:1967by}%
  \BibitemOpen
  \bibfield  {author} {\bibinfo {author} {\bibfnamefont {F.~J.}\ \bibnamefont
  {Ernst}},\ }\href {\doibase 10.1103/PhysRev.168.1415} {\bibfield  {journal}
  {\bibinfo  {journal} {Phys. Rev.}\ }\textbf {\bibinfo {volume} {168}},\
  \bibinfo {pages} {1415} (\bibinfo {year} {1968})}\BibitemShut {NoStop}%
\bibitem [{\citenamefont {Bardeen}(1970)}]{Bardeen:1970}%
  \BibitemOpen
  \bibfield  {author} {\bibinfo {author} {\bibfnamefont {J.~M.}\ \bibnamefont
  {Bardeen}},\ }\href@noop {} {\bibfield  {journal} {\bibinfo  {journal}
  {Astrophys.~J.}\ }\textbf {\bibinfo {volume} {161}},\ \bibinfo {pages} {103}
  (\bibinfo {year} {1970})}\BibitemShut {NoStop}%
\end{thebibliography}%

\end{document}